\documentclass[
 reprint,
superscriptaddress,
 amsmath,amssymb,
 aps,prc,
]{revtex4-2}
\usepackage{multirow}
\usepackage{graphicx}
\usepackage{dcolumn}
\usepackage{bm}
\usepackage{upgreek}
\usepackage[hidelinks]{hyperref}
\usepackage{CJKutf8}
\begin{document}
\preprint{APS/PLACEHOLDER}
\title{Emergence of collectivity in the $0^+_2$ band of $^{92}$Zr}

\author{E.~J.~Williams}
 \email{ewilliams@triumf.ca}
\affiliation{TRIUMF, 4004 Wesbrook Mall, Vancouver, British Columbia, Canada V6T 2A3}

\author{P.~E.~Garrett}
\affiliation{University of Guelph, 50 Stone Rd E, Guelph, Ontario, Canada N1G 2W1}
\affiliation{Department of Physics, University of the Western Cape, P/B X17, 
Bellville, ZA-7535, South Africa}

\author{A.~A.~Avaa}
\affiliation{TRIUMF, 4004 Wesbrook Mall, Vancouver, British Columbia, Canada V6T 2A3}

\author{A.~B.~Garnsworthy}
\affiliation{TRIUMF, 4004 Wesbrook Mall, Vancouver, British Columbia, Canada V6T 2A3}

\author{G.~Hackman}
\affiliation{TRIUMF, 4004 Wesbrook Mall, Vancouver, British Columbia, Canada V6T 2A3}

\author{L.~Mantle}
\affiliation{TRIUMF, 4004 Wesbrook Mall, Vancouver, British Columbia, Canada V6T 2A3}

\author{S.~Murillo~Morales}
\affiliation{TRIUMF, 4004 Wesbrook Mall, Vancouver, British Columbia, Canada V6T 2A3}

\author{J.~R.~Murias}
\affiliation{TRIUMF, 4004 Wesbrook Mall, Vancouver, British Columbia, Canada V6T 2A3}

\author{C.~E.~Svensson}
\affiliation{TRIUMF, 4004 Wesbrook Mall, Vancouver, British Columbia, Canada V6T 2A3}
\affiliation{University of Guelph, 50 Stone Rd E, Guelph, Ontario, Canada N1G 2W1}

\author{R.~Umashankar}
\affiliation{TRIUMF, 4004 Wesbrook Mall, Vancouver, British Columbia, Canada V6T 2A3}
\affiliation{Department of Physics and Astronomy, University of British Columbia, Vancouver, British Columbia, Canada V6T 1Z4}

\author{Y.~Zhu}
\affiliation{TRIUMF, 4004 Wesbrook Mall, Vancouver, British Columbia, Canada V6T 2A3}

\date{\today}

\begin{abstract}
The $2^+_2 \rightarrow 0^+_2$ transition in $^{92}$Zr was observed for the first time in a high-statistics decay study of the 10-day isomer $^{92m}$Nb. Its $E2$ transition strength of 8.9(7) W.u.~is enhanced by 40\% relative to the ground state band, providing evidence for a collective structure built on the $0^+_2$ state which is analogous to the deformed shape-coexisting configuration in $^{94}$Zr.
\end{abstract}

\maketitle

\par The zirconium isotopic chain has proven to be one of the most captivating for nuclear structure studies due to the wide variety of phenomena that it displays. From shell-model seniority structures in $^{90}$Zr (e.g.~\cite{Dey_90Zr}), shape-coexisting structures for the heavier Zr isotopes $^{94,96}$Zr \cite{Chakraborty_94Zr,Kremer_96Zr}, and a quantum phase transition in the shape of the ground state occurring between spherical $N=58$ and well-deformed $N=60$ \cite{Garcia-Ramos_Zr,Gavrielov_Zr,Togashi_Zr}, the Zr isotopes are ideal for studying the evolution of the interplay between microscopic and collective degrees of freedom. A key question that has yet to be fully resolved by experiment is how the collective structures in the heavier Zr isotopes, already apparent in $^{94}$Zr, emerge from the underlying shell-model configurations observed in $^{90}$Zr. A detailed investigation of the intermediate nucleus $^{92}$Zr is needed to answer this question.

The $N=50$ nucleus $^{90}$Zr has a level structure characteristic of a doubly-closed-shell nucleus, with a very high $2^+_1$ energy of 2186 keV, and is one of the few examples of a nucleus with a $J^{\pi}=0^+$ first excited state.  The ground state was assigned as having nearly equal contributions of the $p_{1/2}$ and $g_{9/2}$ proton orbitals, with a wave function $0.8(p_{1/2})^2 +0.6(g_{9/2})^2$, and the $0^+_2$ state involving the orthogonal combination \cite{90Zr_pd}.  The strong mixing of the two configurations is reflected by approximately equal $B(E2;2^+_1\rightarrow 0^+_{1,2})$ values \cite{nds90zr,garrett_90Zr}.  In the $N=52$ nucleus $^{92}$Zr, the $0^+_{1,2}$ states have also been assigned as dominated by the same proton orbitals but with equal mixing (amplitudes of $\approx 0.7$) \cite{Zr_d3He,Zr_3Hed,Saha}.  The $2^+_1$ state, on the other hand, is located at a much lower energy than in $^{90}$Zr, and is mostly a neutron excitation involving the $d_{5/2}$ orbital \cite{Sieja_Zr_SM}, as is the $4^+_1$ state.  When we reach the $N=54$ nucleus $^{94}$Zr, we would expect the $0^+_1$, $2^+_1$, and $4^+_1$ states to have properties very similar to those of $^{92}$Zr since they would also be dominated by the same proton and neutron configurations. The $2^+_1$ and $4^+_1$ excitation energies are indeed nearly identical. However, there is a significant drop in the $B(E2;4^+_1\rightarrow 2^+_1)$ value in $^{94}$Zr compared to $^{92}$Zr \cite{Mach_Zr}.

The $0^+_2$ state in $^{94}$Zr was assigned by Chakraborty {\it et al.}\ \cite{Chakraborty_94Zr} as the band head of a deformed shape-coexisting configuration based on the observation of significantly enhanced $B(E2;2^+_2\rightarrow 0^+_2)$ and $B(E2;4^+_2\rightarrow 2^+_2)$ values of $19(2)$ and $34^{+10}_{-17}$ W.u., respectively (which were recently remeasured to be 18.8(9) and 45(3) W.u.~\cite{Naomi_94Zr}).  The assignment of a deformed band was in qualitative agreement with large-scale shell model calculations \cite{Sieja_Zr_SM} that predicted two-particle two-hole proton excitations from the $p_{1/2}$ and $p_{3/2}$ orbitals to the $g_{9/2}$ orbital. A later Coulomb excitation experiment determined that the $0^+_2$ state corresponds to a deformed triaxial-oblate shape \cite{Naomi_94Zr}.  A deformed shape-coexistence structure was also invoked for the $0^+_2$ state in $^{96}$Zr, based on a deduced $B(E2;2^+_2\rightarrow 0^+_2)$ value of 36(11) W.u.~\cite{Kremer_96Zr}. The properties of the $2^+_2$ state in $^{96}$Zr were recently remeasured, yielding $B(E2;2^+_2\rightarrow 0^+_2)=38.9(57)$ W.u.~\cite{Zielinska_96Zr}. This higher precision measurement reveals a substantial increase in the quadrupole collectivity of the $0^+_2$ band compared to $^{94}$Zr despite the {\it increase} in the $E(2^+_2)-E(0^+_2)$ energy spacing (371 keV and 644 keV for $^{94}$Zr and $^{96}$Zr, respectively). Combined with observations from transfer reactions, it was suggested \cite{Zielinska_96Zr} that $^{94,96}$Zr may possess a triple-shape-coexistence scenario as predicted by both the Monte Carlo shell model \cite{Togashi_Zr} and the complex excited Vampir calculations \cite{Petrovici_96Zr}.

Given these observations, it is natural to ask if a deformed configuration also exists in $^{92}$Zr. Sheline {\it et al.}\ \cite{Sheline_1972} have postulated the appearance of deformed structures throughout the Zr isotopes, and in particular suggested the sequence $0^+_2$, $2^+_2$, and $4^+_2$ in $^{92}$Zr. In an investigation of high-spin states in $^{92}$Zr, Sugawara {\it et al.}\ \cite{Sugawara_92Zr} used rotational expressions to extrapolate the energies of the high-spin states to spin 0, finding reasonable agreement with the energies of the $0^+_2$ and $2^+_2$ states. They further predicted that the $B(E2;2^+_2\rightarrow 0^+_2)$ value should be 18 W.u., similar to the deformed configuration in $^{94}$Zr. More recently, Gavrielov and co-workers \cite{Gavrielov_Zr} described the Zr isotope chain as undergoing an intertwined quantum phase transition within the interacting boson model configuration mixing (IBM-CM) framework. Two configurations were considered in their work: a normal configuration with no proton excitations above the $Z=40$ closed subshell that is approximately spherical throughout the chain, and a two-particle two-hole intruder configuration that undergoes a smooth transition from spherical to deformed with increasing neutron number beginning with $^{92}$Zr. These configurations cross with the $2^+_1$ state transitioning from the normal to the intruder configuration between $N=56$ and $N=58$, and the $0^+_1$ state undergoing the same transition between $N=58$ and $N=60$. It was assumed that the intruder configuration in $^{92}$Zr was based on the $0^+_2$, $2^+_2$, $4^+_2$, and $6^+_1$ states, and the chosen parameters for the intruder Hamiltonian reproduced the energies reasonably well (see Fig.~4 of Ref.~\cite{Gavrielov_Zr}). We emphasize that Gavrielov {\it et al.}~do not suggest that the intruder configuration in $^{92}$Zr is deformed, but rather has a spherical nature while remaining distinct from the ground state configuration. 

\par In the current work, we have used the $\beta^+$/EC decay of $^{92m}$Nb ($t_{1/2} \approx 10$ days) to directly populate low-lying $2^+$ states in $^{92}$Zr and establish the branching ratio of the $2^+_2\rightarrow 0^+_2$ transition via $\gamma$-ray spectroscopy. Our measurement reveals a substantial enhancement of the $2^+_2\rightarrow 0^+_2$ transition rate compared to transitions in the ground state band. With this result, an important step has been realized in understanding the emergence of collectivity originating from $^{90}$Zr towards the deformed shape-coexisting configurations in the heavier Zr isotopes. We have also obtained improved half-life and branching ratio measurements for $^{92m}$Nb decay.

\par Two sources of $^{92m}$Nb were prepared by irradiating natural niobium foils (99.8\% purity) purchased from Fisher Scientific with 13 MeV protons from the TR13 medical cyclotron at TRIUMF \cite{tr13,tr13_targetsys}, at 5 $\upmu$A for 15 minutes.  Most of the activity observed after each irradiation was due to the 6.85-hour isomer of $^{93}$Mo, which was allowed to decay for at least 72 hours before each measurement. Data were collected using the GRIFFIN array \cite{griffin2019} populated with 16 high-purity germanium clover detectors. The first source was measured for 21 days with the front face of the GRIFFIN clover detectors positioned 110 mm from the source, and the second source was measured for 17 days with the clover detectors  positioned 145 mm from the source and Compton-suppression shields pulled forward to veto Compton scattering between the clover detectors. Energy summing (addback) was used for all time-coincident ($\Delta t \leq 200$ ns) hits in crystals positioned $\leq 100$ mm apart, restricted to crystals from a single clover detector for the data with forward-positioned suppression shields.  Summing corrections were applied using the 180$^{\circ}$ coincidence procedure of Ref.~\cite{griffin2019}. Energy and absolute efficiency calibrations were obtained using $^{56,60}$Co, $^{133}$Ba, and $^{152}$Eu sources.

\par
\begin{figure}
\begin{center}
\includegraphics[width=\columnwidth]{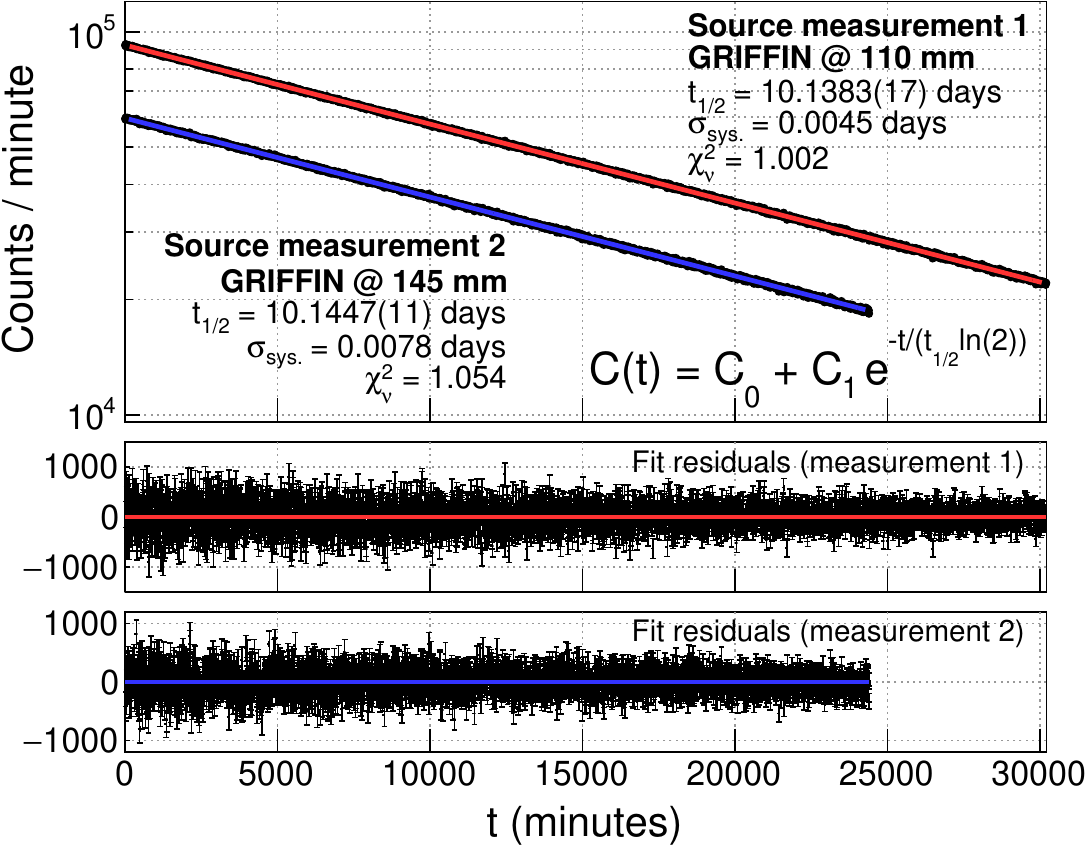}
\end{center}
\vspace{-0.3cm}
\caption{Decay curves of $^{92m}$Nb, from Compton background subtracted 934.45(6) keV photopeak data.}
\label{fig:decaycurve}
\end{figure}

 The half-life of $^{92m}$Nb was determined by fitting the decay curve of the 934.45(6) keV photopeak after subtracting the Compton-continuum background resulting from higher energy transitions. Decay curve fits are shown in Figure \ref{fig:decaycurve}. Various modified fit conditions were used to test for systematic uncertainties: changing the energy window used for background subtraction, scaling the subtracted background by $\pm 10$\%, excluding the first or last 10\% of the data, and/or adding additional fit terms corresponding to short-lived ($t_{1/2} \leq 9$ days) and long-lived ($t_{1/2} \geq 11$ days) contaminants. We found that the 2\textsuperscript{nd} source measurement fit in particular was sensitive to the background subtraction and data exclusion procedure, likely due to intermittent room background from nearby experiments which were running at the same time as this measurement. The best fit $^{92m}$Nb half-life was $10.1383 \pm 0.0017_{stat.} \pm 0.0045_{sys.}$ days for the 1\textsuperscript{st} source measurement and $10.1447 \pm 0.0011_{stat.} \pm 0.0078_{sys.}$ days for the 2\textsuperscript{nd} source measurement. Since the systematic errors are dominant, we adopt the unweighted average of $10.142(5)$ days, where the uncertainty is inflated to match the systematic error of the 1\textsuperscript{st} measurement.

\begin{figure}
\begin{center}
\includegraphics[width=\columnwidth]{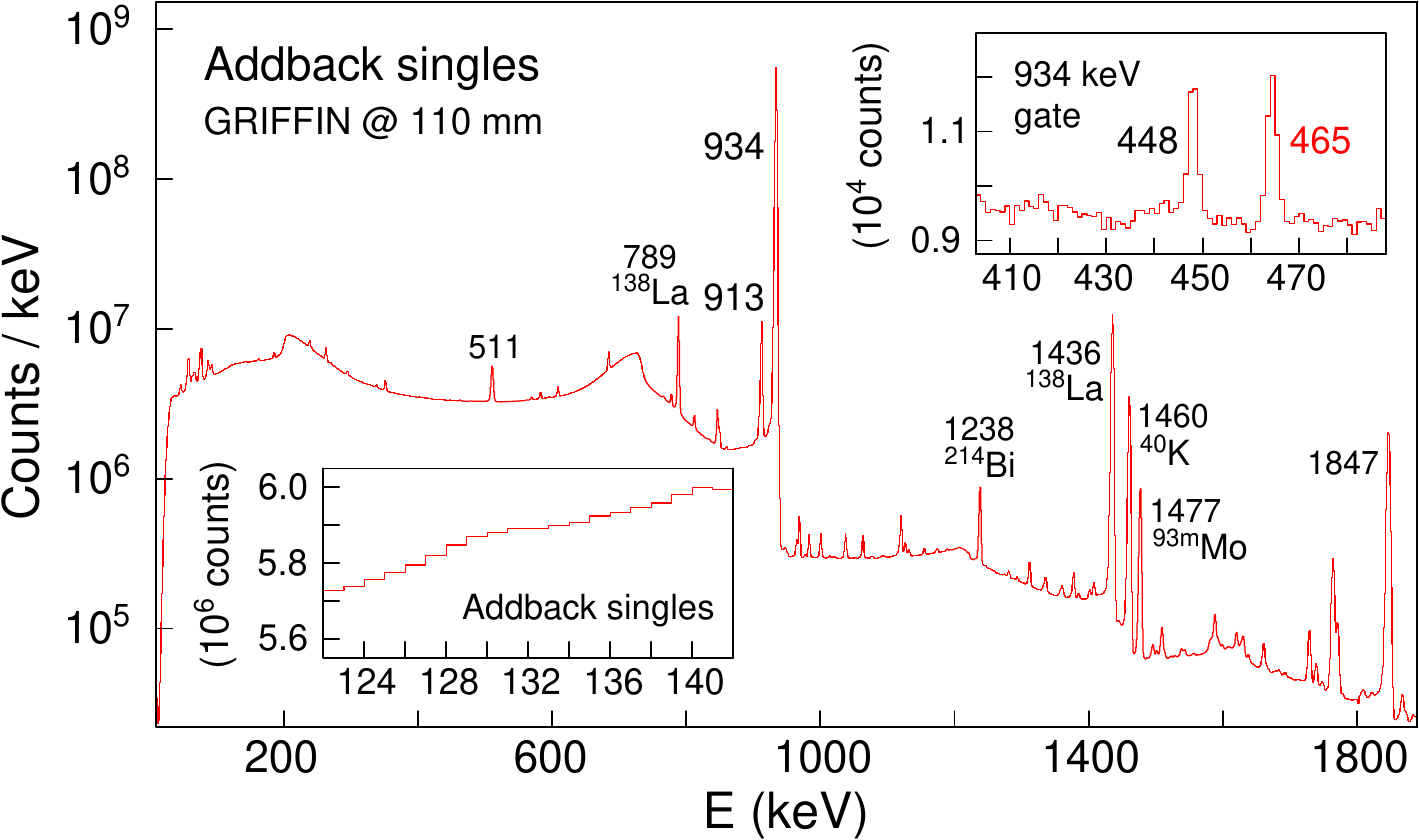}
\end{center}
\caption{Energy spectrum obtained using GRIFFIN, with insets showing the region of the unobserved 135.5 keV ground-state transition in $^{92}$Nb, and the $2^+_2 \rightarrow 0^+_2 \rightarrow 2^+_1$ cascade in $^{92}$Zr. The 464.61(9) keV transition is newly observed.}
\label{fig:spectra}
\end{figure}

\par
\begin{figure}
\begin{center}
\includegraphics[width=\columnwidth]{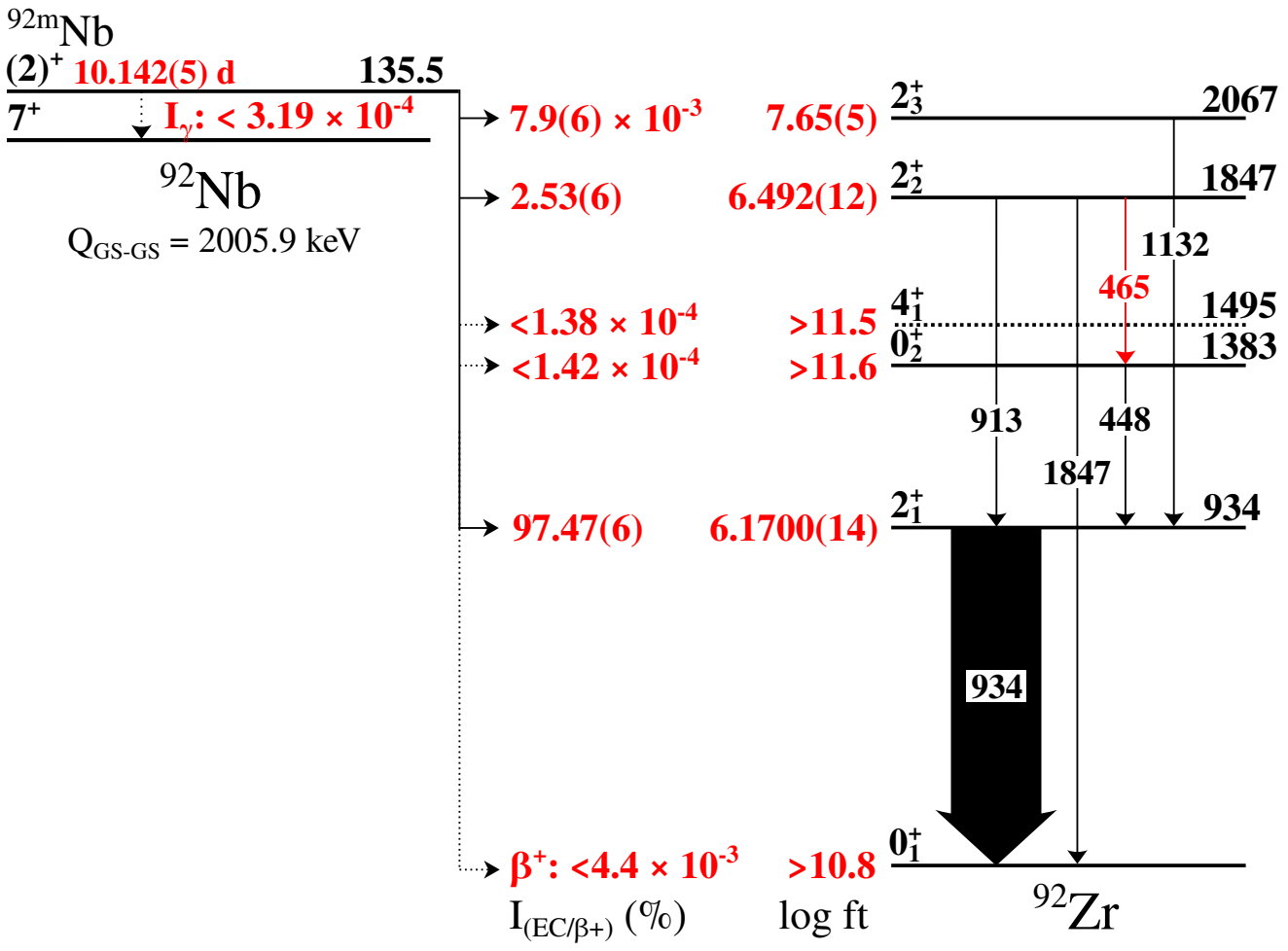}
\end{center}
\vspace{-0.3cm}
\caption{Decay scheme of $^{92m}$Nb: values in red are from this work, all other values are from Ref.~\cite{nds92}. Limits are reported at 95\% ($2\sigma$) confidence.}
\label{fig:decayscheme}
\end{figure}

 Figure \ref{fig:spectra} shows a $\gamma$-ray energy spectrum. Natural background lines from $^{40}$K (1460 keV), $^{138}$La (789, 1436 keV, from nearby LaBr$_3$ scintillators not used in this measurement), and thorium/uranium decay chains (eg.~$^{214}$Bi, $^{228}$Ac) were present in the data. Relative intensities of transitions following $^{92m}$Nb decay are reported in Table \ref{tab:92mNb_decay_table}, corrected for internal conversion using coefficients obtained from the BrIcc code \cite{2008Ki07}. For unobserved transitions, the $1\sigma$ upper limit in the number of photopeak counts was taken as $\sqrt{2B}$, where $B$ is the number of background counts in a window centered on the expected peak energy, with a width of twice the expected full width at half-maximum of the peak as determined from the calibration data. 
Figure \ref{fig:decayscheme} shows the overall decay scheme deduced, with decay intensities and log $ft$ values derived from the values of Table \ref{tab:92mNb_decay_table}.

\par We sought evidence of the direct population of the $^{92}$Zr ground state by counting positron annihilation events ($180^{\circ}$ coincident 511 keV $\gamma$ rays), but found none for this decay branch after accounting for background from other positron emitters produced by reactions with small ($\leq 0.2$\%) impurities in our target foil ($^{48}$V, $^{52}$Mn, $^{56}$Co, $^{89}$Zr, $^{90}$Nb). The resulting limit for the $\beta^+$ branch to the ground state was $< 4.4 \times 10^{-3}$~\% ($2\sigma$ confidence), corresponding to a log$ft$ lower limit of $10.8$, consistent with the expected 2\textsuperscript{nd} forbidden character of this transition. The unobserved decay branches to the $0^+_2$ and $4^+_1$ levels in $^{92}$Zr were also consistent with 2\textsuperscript{nd} forbidden transitions (log$ft$ $>11.5$). These observations support a spin-parity assignment of $2^+$ for the $^{92m}$Nb parent. Combining our results with previous $^{93}$Nb($d,t$) and $^{93}$Nb($p,d$) data \cite{bhatia1971,rumore1984}, the only other possible spin-parity is $3^+$, however this would likely result in a large decay branch to the $7^+$ ground state of $^{92}$Nb, which we do not observe. Based on our upper limit for the $\gamma$ decay branch ($3.19 \times 10^{-4}$~\%), a $3^+$ assignment for $^{92m}$Nb would result in $B(E4; 3^+_1 \rightarrow 7^+_1) < 9.38 \times 10^{-5}$~W.u. This would be significantly hindered compared to typical \emph{E4} strengths in this mass region (0.1 -- 10~W.u.), so it seems likely that the $IT$ decay proceeds by an as-yet unobserved $2^+ \rightarrow 7^+$ $M5$ transition instead.

\par
\begin{figure}
\begin{center}
\includegraphics[width=\columnwidth]{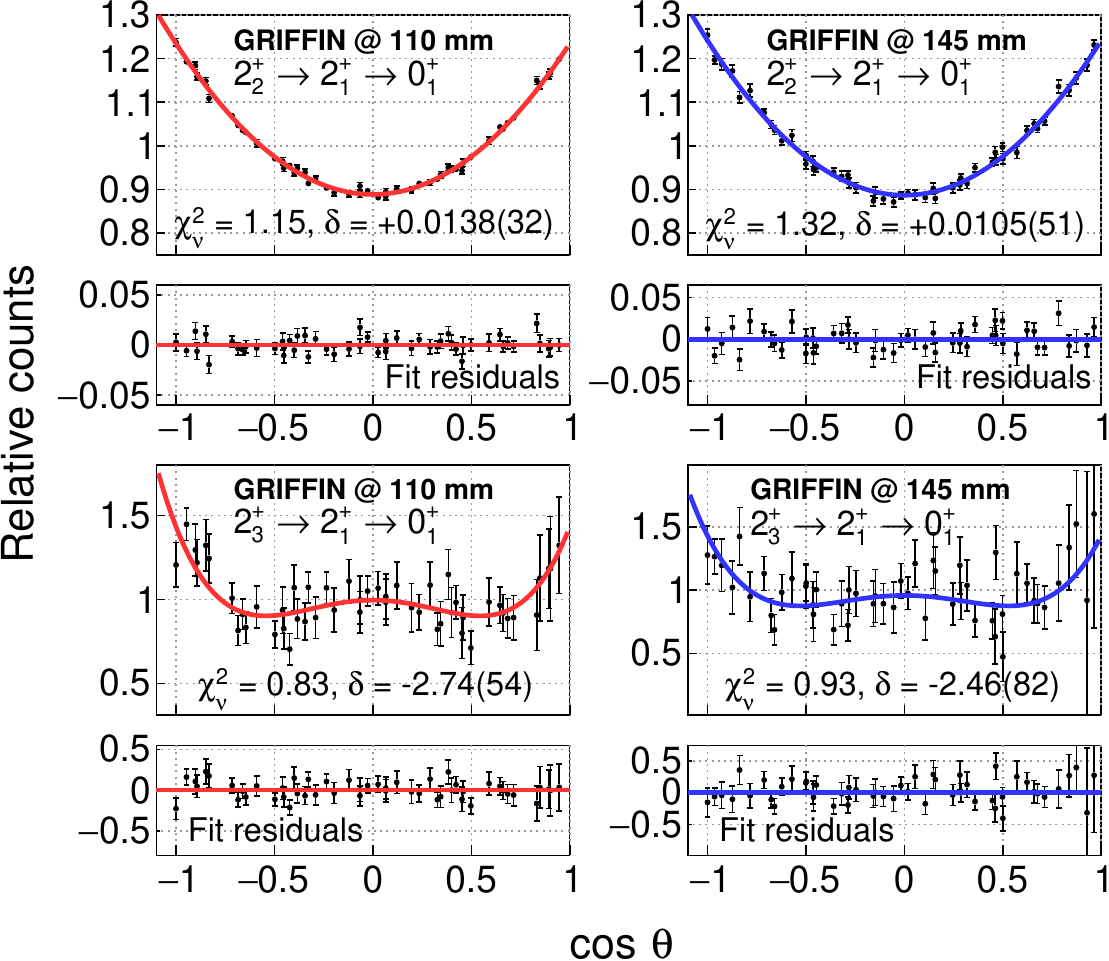}
\end{center}
\vspace{-0.3cm}
\caption{Angular correlations of $2^+ \rightarrow 2^+ \rightarrow 0^+$ cascades in $^{92}$Zr, compared to theoretical functions using the best-fit $\delta$ value for each $2^+ \rightarrow 2^+$ transition.}
\label{fig:angcorr}
\end{figure}

 Multipole mixing ratios for $E2/M1$ transitions were determined using $\gamma-\gamma$ angular correlations, with the $2^+_2 \rightarrow 2^+_1 \rightarrow 0^+_1$ cascade used to determine the mixing ratio for the 912.68(9) keV $2^+_2 \rightarrow 2^+_1$ transition, and the $2^+_3 \rightarrow 2^+_1 \rightarrow 0^+_1$ cascade used to determine the mixing ratio for the 1132.20(7) keV $2^+_3 \rightarrow 2^+_1$ transition. Angular correlation data were normalized using the event-mixing technique described in Ref.~\cite{smith_gamma-gamma_2019}, and compared to theoretical distributions corrected for finite detector size effects using the energy-dependent $\beta$ and $\gamma$ parametrization also described in Ref.~\cite{smith_gamma-gamma_2019}. A $\chi^2$ minimization was performed to obtain the best-fit mixing ratio $\delta$, with the statistical uncertainty determined using the $\chi^2_{min} + 1$ bounds.  The $2^+_2 \rightarrow 2^+_1 \rightarrow 0^+_1$ angular correlation analysis was corrected for the presence of significant background from Compton scattering of the $2^+_2 \rightarrow 0^+_1$ 1847.34(6) keV $\gamma$ ray, by subtracting the angular correlation obtained using offset $\gamma$-ray energies that sum to 1847 keV. Angular correlation fits are shown in Figure \ref{fig:angcorr}. For the $2^+_2 \rightarrow 2^+_1$ transition, mixing ratios $\delta = +0.0138(32), +0.0105(51)$ were obtained for each source dataset (based on the Krane-Steffen convention \cite{krane_determination_1970}), yielding a weighted average of $+0.013(3)$. For the $2^+_3 \rightarrow 2^+_1$ transition, mixing ratios $\delta = -2.74(54), -2.46(82)$ were obtained, yielding a weighted average of $-2.7(5)$. These mixing ratios are in agreement with the adopted values of Ref.~\cite{nds92}. Previously, the sign of the $2^+_2 \rightarrow 2^+_1$ mixing ratio was ambiguous due to significant disagreement between the measurements of Refs.~\cite{bunker_decay_1962,Fransen_92Zr}. Our data strongly support a positive value and result in a non-zero $B(E2; 2^+_2 \rightarrow 2^+_1)$ of $0.05(2)$ W.u.~($0.050^{+0.080}_{-0.042}$ W.u.~at 3$\sigma$ confidence).

\begin{table*}
\caption{Transitions following $^{92m}$Nb decay with their relative intensities and $E2/M1$ mixing ratios $\delta$ determined from each source measurement. Transition strengths $B(\lambda L; J^{\pi}_i \rightarrow J^{\pi}_f)$ were derived using lifetimes from Ref.~\cite{peters_level_2013} where available and Ref.~\cite{nds92} otherwise. Previously unmeasured values are in bold. Upper limits are reported at 95\% ($2\sigma$) confidence.}
\centering
\begin{ruledtabular}
\begin{tabular}{lccccccccr}
\multirow{3}{*}{$E_{\gamma}$ (keV)} & \multirow{3}{*}{$J^{\pi}_i$} & \multirow{3}{*}{$J^{\pi}_f$} & \multicolumn{3}{c}{Relative intensity (\%)} & \multirow{3}{*}{$\delta$} & \multirow{3}{*}{$\lambda L$} & \multicolumn{2}{c}{$B(\lambda L; J^{\pi}_i \rightarrow J^{\pi}_f)$ (W.u.)} \\ \cline{4-6} \cline{9-10}
\rule{0pt}{2.5ex} &  &  & GRIFFIN  & GRIFFIN & Weighted &  & & \multirow{2}{*}{This work} & \multirow{2}{*}{\shortstack[r]{IBM-CM\\calc.~\cite{Gavrielov_Zr}}} \\ 
\rule{0pt}{2.5ex} &  &  & @ 110 mm  & @ 145 mm & average &  & & &  \\ \cline{1-8} \cline{9-10}
\rule{0pt}{2.5ex}(135.5)$^{a}$ & $(2)^+$ & $7^+$ & $\mathbf{< 4.58 \times 10^{-4}}$ $^b$ & $\mathbf{< 4.44 \times 10^{-4}}$ $^b$ & $\mathbf{< 3.19 \times 10^{-4}}$ $^b$ &  & $(M5)$ & $\mathbf{< 7.49 \times 10^{4}}$ & \\ 
(351.8) & $2^+_2$ & $4^+_1$ & $\mathbf{< 1.55 \times 10^{-4}}$ & $\mathbf{< 2.33 \times 10^{-4}}$ & $\mathbf{< 1.29 \times 10^{-4}}$ &  & $E2$ & $\mathbf{< 2.77}$ $^c$ & \\
448.19(9)               & $0^+_2$ & $2^+_1$ & $\mathbf{1.75(14) \times 10^{-3}}$ & $\mathbf{1.84(15) \times 10^{-3}}$ & $\mathbf{1.79(10) \times 10^{-3}}$ &  & $E2$ & $14.4(5)$ $^d$ & 1.8 \\
$\mathbf{464.61(9)}$     & $2^+_2$ & $0^+_2$ & $\mathbf{1.84(14) \times 10^{-3}}$ & $\mathbf{1.78(15) \times 10^{-3}}$ & $\mathbf{1.81(10) \times 10^{-3}}$ &  & $E2$ & $\mathbf{8.9(7)}$ $^c$ & 17\\
(560.9)                  & $4^+_1$ & $2^+_1$ & $< 1.60 \times 10^{-4}$   & $< 2.96 \times 10^{-4}$ & $< 1.41 \times 10^{-4}$ &  & $E2$ & $4.05(12)$ $^d$ & \\
(571.3)                  & $2^+_3$ & $4^+_1$ & $< 1.64 \times 10^{-4}$ & $< 3.00 \times 10^{-4}$ & $< 1.44 \times 10^{-4}$ &  & $E2$ & $4^{+3}_{-2}$ $^e$ & \\
(683.8)                  & $2^+_3$ & $0^+_2$ & $\mathbf{< 2.93 \times 10^{-4}}$ & $\mathbf{< 4.78 \times 10^{-4}}$ & $\mathbf{< 2.50 \times 10^{-4}}$ &  & $E2$ & $\mathbf{< 14.56}$ $^c$ & \\
\rule{0pt}{3.0ex}\multirow{2}{*}{912.68(9)}                & \multirow{2}{*}{$2^+_2$} & \multirow{2}{*}{$2^+_1$} & \multirow{2}{*}{$1.78(10)$} & \multirow{2}{*}{$1.776(61)$} & \multirow{2}{*}{$1.78(5)$} & \multirow{2}{*}{+0.013(3)} & $M1$ & 0.239(13) $^c$ &  \\
                         & & & & & & & $E2$ & 0.05(2) $^c$ & 0.095 \\
\rule{0pt}{3.0ex}934.45(6)                & $2^+_1$ & $0^+_1$ & 100.0(23)                   & 100.0(14) & 100.0(12) &  & $E2$ & $6.4^{+0.6}_{-0.5}$ $^d$ & 1.1 \\

\rule{0pt}{3.0ex}\multirow{2}{*}{1132.20(7)}              & \multirow{2}{*}{$2^+_3$} & \multirow{2}{*}{$2^+_1$} & \multirow{2}{*}{$7.93(58) \times 10^{-3}$}   & \multirow{2}{*}{$7.48(73) \times 10^{-3}$} & \multirow{2}{*}{$7.8(5) \times 10^{-3}$} & \multirow{2}{*}{-2.7(5)} & $M1$ & $0.003^{+0.002}_{-0.001}$ $^e$ &  \\
                         & & & & & & & $E2$ & $19^{+9}_{-5}$ $^e$ & 0.87\\
                         
\rule{0pt}{3.0ex}1847.34(6)              & $2^+_2$ & $0^+_1$ & $0.779(29)$   & $0.720(62)$ & $0.77(3)$ &  & $E2$ & $3.8(2)$ $^c$ & 0.50 \\
(2066.7)                 & $2^+_3$ & $0^+_1$ & $< 9.46 \times 10^{-5}$   & $< 1.74 \times 10^{-4}$ & $< 8.31 \times 10^{-5}$ &  & $E2$ & $6^{+3}_{-2} \times 10^{-3}$ $^e$ & $4 \times 10^{-5}$ \\
\end{tabular}
\end{ruledtabular}
\label{tab:92mNb_decay_table}
{\footnotesize \rule{0pt}{2.5ex}$^{a}$Transition in $^{92}$Nb, all other transitions are in $^{92}$Zr. $^b$Not corrected for internal conversion. $^c$Using lifetime data from Ref.~\cite{peters_level_2013}. $^d$Using lifetime and branching data from Ref.~\cite{nds92}. $^e$Using lifetime data from Ref.~\cite{peters_level_2013} and branching data from Ref.~\cite{nds92}.}
\end{table*}

\par A previously unobserved line at 464.61(9) keV was assigned as the $2^+_2 \rightarrow 0^+_2$ transition in $^{92}$Zr based on its time-coincidence with the 448.19(9) keV and 934.45(6) keV transitions. Using the $2^+_2$ level lifetime measured by Peters {\it et al.}~\cite{peters_level_2013} and the $2^+_2 \rightarrow 0^+_2$ branch of $7.1(4) \times 10^{-2}$~\% determined in this work, we obtain a $B(E2;2^+_2 \rightarrow 0^+_2)$ value of $8.9(7)$ W.u. This is approximately 40\% larger than the $2^+_1 \rightarrow 0^+_1$ value of $6.4^{+0.6}_{-0.5}$ W.u., indicating a slight enhancement of collectivity in the $0^+_2$ band. Gavrielov {\it et al.}\ \cite{Gavrielov_Zr} suggested that the $4^+_2$ level is also associated with the configuration based on the $0^+_2$ state. Very recently, the existence of the 551 keV $4^+_2\rightarrow 2^+_2$ transition was reported \cite{Urban}. A $B(E2;4^+_2\rightarrow 2^+_2)$ value of $23^{+8}_{-7}$ W.u.\ is obtained using the evaluated lifetime \cite{nds92}, indicating a collective enhancement and firmly placing the $4^+_2$ level as a member of the $0^+_2$ band (the $4^+_2$ lifetime from Ref.~\cite{peters_level_2013} is significantly shorter, leading to $44(14)$ W.u.).

\begin{figure}[t!]
\begin{center}
\includegraphics[width=\columnwidth]{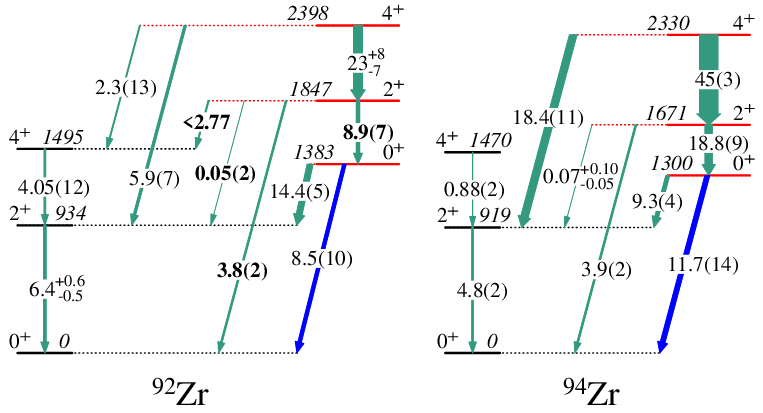}
\end{center}
\vspace{-0.3cm}
\caption{\label{fig:92-94Zr-comp}Partial level schemes of $^{92}$Zr and $^{94}$Zr: transitions are labelled with $B(E2)$ values in W.u.\ (green arrows) or $10^3\times \rho^2(E0)$ values (blue arrows). Bolded values are from this work, all other values are from Refs.~\cite{Mach_Zr,Naomi_94Zr,nds92,Urban,KibediE0}.}
\end{figure}

Figure \ref{fig:92-94Zr-comp} compares the level schemes of the low energy states in $^{92}$Zr and $^{94}$Zr. Focusing on the states assigned to the $0^+_2$ configuration, there are many common features.  Firstly, for both isotopes, the $0^+_2$ state has an enhanced transition to the $2^+_1$ state and a moderate $\rho^2(E0)$ value for the $0^+_2\rightarrow 0^+_1$ transition.  This latter observation implies that there must be differences in the mean-square charge radii $\langle r^2\rangle$ as well as mixing between the underlying configurations involved in the $0^+_1$ and $0^+_2$ states \cite{Wood_E0}. The $2^+_2$ states also have moderate enhancements in the $B(E2;2^+_2\rightarrow 0^+_2)$ value (8.9(7) W.u.\ and 18.8(9) W.u.\ for $^{92}$Zr and $^{94}$Zr, respectively), as well as weakly collective transitions to the ground state (3.8(2) W.u.\ and 3.9(2) W.u., respectively). The transitions to the $2^+_1$ states are dominated by the $M1$ components, with extremely weak $E2$ strengths and $B(M1;2^+_2\rightarrow 2^+_1)=0.239(13)$ W.u.\ and 0.153(12) W.u., respectively. The measured $g$ factors for the $2^+_2$ states are $+0.76(50)$ and $+0.88(27)$, respectively, opposite in sign to those of the $2^+_1$ states, and indicative of a proton dominance in the wave functions \cite{Werner_gfact}. Finally, the $4^+_2$ states in both isotopes show a moderate enhancement in their $B(E2;4^+_2\rightarrow 2^+_1$) values and a significantly larger enhancement in their $B(E2;4^+_2\rightarrow 2^+_2$) values, clearly indicating a band structure.

Both $^{92}$Zr and $^{94}$Zr have been studied by a variety of nucleon-transfer reactions.  In the $^{90}$Zr($t,p$) reaction \cite{Flynn_tp}, the $0^+_2$ state was populated with 5\% of the strength of the ground state, and in $^{94}$Zr the corresponding value was 4\%.  A substantial difference in the population of the $2^+_2$ states was observed, however, with a weaker population in $^{92}$Zr (9\%) but strong in $^{94}$Zr (32\%) relative to the ground state cross sections \cite{Flynn_tp} indicating a significant change in the microscopic composition.  Using the ($p,t$) reaction, consistent populations of the $0^+_2$ states relative to the ground states were found (3\% for $^{92}$Zr and 5\% for $^{94}$Zr, respectively), but the opposite pattern compared with the ($t,p$) reaction for the $2^+_2$ state with $^{92}$Zr having the greater population (18\% relative to the ground state cross section) than in $^{94}$Zr (4\%) \cite{Ball_pt}.  The two-proton transfer reactions $^{94,96}$Mo($^6$Li,$^8$B) found similar populations of the $0^+_2$ states (20\% and 24\% for $^{92}$Zr and $^{94}$Zr, respectively) \cite{Tickle_Zr}.  These latter results point to enhanced proton pair correlations in the $0^+_2$ state.  Finally, $\alpha$-particle pickup reactions performed on the Mo isotopes populated the $0^+_2$ state in $^{92}$Zr with approximately 10\% of the ground state strength, and 8\% in $^{94}$Zr \cite{Saha,VanDenBerg}.  

Given these observations, we suggest that the $0^+_2$ state in $^{92}$Zr has strong structural similarities to the $0^+_2$ state in $^{94}$Zr.  This is in agreement with the IBM-CM calculations of Gavrielov {\it et al.}\ \cite{Gavrielov_Zr}, which predicted a $B(E2;2_2^+\rightarrow 0^+_2)$ value of $17$ W.u.\ compared to our result of 8.9(7) W.u.  As these calculations were not fine-tuned to reproduce any individual isotope, but rather to reflect general trends across the isotopic chain, we do not find the factor of two difference between our result and the theory prediction to be a significant disagreement.  The IBM-CM calculations predict a $B(E2;4^+_2\rightarrow 2^+_2)$ value of $21$ W.u.\ in excellent agreement with the experimental data, but with a weak transition to the $2^+_1$ state (0.97 W.u.)~compared to the experimental result (5.9(7) W.u.)~which indicates that there is a greater mixing of the configurations than accounted for in the calculations. The MCSM calculations of Togashi {\it et al.}\ \cite{Togashi_Zr}, with the T-plots shown in Fig.~52 of Ref.~\cite{Leoni_PPNP}, also point to a structural similarity of the $0^+_2$ configurations in $^{92}$Zr and $^{94}$Zr, albeit with a greater mixing of the spherical and oblate configurations in $^{92}$Zr.

\par In summary, we have observed the $2^+_2 \rightarrow 0^+_2$ transition in $^{92}$Zr for the first time, as part of a detailed decay study of $^{92m}$Nb. Its extracted branching ratio leads to a $B(E2;2^+_2\rightarrow 0^+_2)$ value of $8.9(7)$ W.u., indicating a collective enhancement. This result permits the firm assignment of the $2^+_2$ state as a member of a collective structure built on the $0^+_2$ level. Based on the strong similarities of the decay patterns and transfer populations, we suggest that the underlying configuration of the $0^+_2$ state in $^{92}$Zr matches that of the $0^+_2$ state in $^{94}$Zr, which has recently been shown to have a triaxial-oblate shape.  In order to firmly establish this scenario, detailed Coulomb excitation studies should be performed to extract the quadrupole moments and quadrupole invariants in a model-independent way from which the intrinsic shapes can be inferred.

\medskip

\begin{acknowledgments}
\par We would like to thank the TR13 operators for performing the proton irradiations and S.~Georges for design and fabrication of the source holder. This work was supported by the Natural Sciences and Engineering Research Council of Canada, the Canadian Foundation for Innovation and the British Columbia Knowledge Development Fund. TRIUMF receives federal funding through a contribution agreement with the National Research Council of Canada.
\end{acknowledgments}


\begin{thebibliography}{40}%
\makeatletter
\providecommand \@ifxundefined [1]{%
 \@ifx{#1\undefined}
}%
\providecommand \@ifnum [1]{%
 \ifnum #1\expandafter \@firstoftwo
 \else \expandafter \@secondoftwo
 \fi
}%
\providecommand \@ifx [1]{%
 \ifx #1\expandafter \@firstoftwo
 \else \expandafter \@secondoftwo
 \fi
}%
\providecommand \natexlab [1]{#1}%
\providecommand \enquote  [1]{``#1''}%
\providecommand \bibnamefont  [1]{#1}%
\providecommand \bibfnamefont [1]{#1}%
\providecommand \citenamefont [1]{#1}%
\providecommand \href@noop [0]{\@secondoftwo}%
\providecommand \href [0]{\begingroup \@sanitize@url \@href}%
\providecommand \@href[1]{\@@startlink{#1}\@@href}%
\providecommand \@@href[1]{\endgroup#1\@@endlink}%
\providecommand \@sanitize@url [0]{\catcode `\\12\catcode `\$12\catcode
  `\&12\catcode `\#12\catcode `\^12\catcode `\_12\catcode `\%12\relax}%
\providecommand \@@startlink[1]{}%
\providecommand \@@endlink[0]{}%
\providecommand \url  [0]{\begingroup\@sanitize@url \@url }%
\providecommand \@url [1]{\endgroup\@href {#1}{\urlprefix }}%
\providecommand \urlprefix  [0]{URL }%
\providecommand \Eprint [0]{\href }%
\providecommand \doibase [0]{https://doi.org/}%
\providecommand \selectlanguage [0]{\@gobble}%
\providecommand \bibinfo  [0]{\@secondoftwo}%
\providecommand \bibfield  [0]{\@secondoftwo}%
\providecommand \translation [1]{[#1]}%
\providecommand \BibitemOpen [0]{}%
\providecommand \bibitemStop [0]{}%
\providecommand \bibitemNoStop [0]{.\EOS\space}%
\providecommand \EOS [0]{\spacefactor3000\relax}%
\providecommand \BibitemShut  [1]{\csname bibitem#1\endcsname}%
\let\auto@bib@innerbib\@empty
\bibitem [{\citenamefont {Dey}\ \emph {et~al.}(2022)\citenamefont {Dey},
  \citenamefont {Negi}, \citenamefont {Palit}, \citenamefont {Srivastava},
  \citenamefont {Laskar}, \citenamefont {Das}, \citenamefont {Babra},
  \citenamefont {Bhattacharya}, \citenamefont {Das}, \citenamefont {Devi},
  \citenamefont {Gala}, \citenamefont {Garg}, \citenamefont {Ghugre},
  \citenamefont {Ideguchi}, \citenamefont {Kumar}, \citenamefont {Kundu},
  \citenamefont {Mukherjee}, \citenamefont {Muralithar}, \citenamefont {Nag},
  \citenamefont {Nandi}, \citenamefont {Neelam}, \citenamefont {Raja},
  \citenamefont {Raut}, \citenamefont {Santra}, \citenamefont {Sharma},
  \citenamefont {Sihotra}, \citenamefont {Singh}, \citenamefont {Singh},\ and\
  \citenamefont {Trivedi}}]{Dey_90Zr}%
  \BibitemOpen
  \bibfield  {author} {\bibinfo {author} {\bibfnamefont {P.}~\bibnamefont
  {Dey}}, \bibinfo {author} {\bibfnamefont {D.}~\bibnamefont {Negi}}, \bibinfo
  {author} {\bibfnamefont {R.}~\bibnamefont {Palit}}, \bibinfo {author}
  {\bibfnamefont {P.~C.}\ \bibnamefont {Srivastava}}, \bibinfo {author}
  {\bibfnamefont {M.~S.~R.}\ \bibnamefont {Laskar}}, \bibinfo {author}
  {\bibfnamefont {B.}~\bibnamefont {Das}}, \bibinfo {author} {\bibfnamefont
  {F.~S.}\ \bibnamefont {Babra}}, \bibinfo {author} {\bibfnamefont
  {S.}~\bibnamefont {Bhattacharya}}, \bibinfo {author} {\bibfnamefont
  {B.}~\bibnamefont {Das}}, \bibinfo {author} {\bibfnamefont {K.~R.}\
  \bibnamefont {Devi}}, \bibinfo {author} {\bibfnamefont {R.}~\bibnamefont
  {Gala}}, \bibinfo {author} {\bibfnamefont {U.}~\bibnamefont {Garg}}, \bibinfo
  {author} {\bibfnamefont {S.~S.}\ \bibnamefont {Ghugre}}, \bibinfo {author}
  {\bibfnamefont {E.}~\bibnamefont {Ideguchi}}, \bibinfo {author}
  {\bibfnamefont {S.}~\bibnamefont {Kumar}}, \bibinfo {author} {\bibfnamefont
  {A.}~\bibnamefont {Kundu}}, \bibinfo {author} {\bibfnamefont
  {G.}~\bibnamefont {Mukherjee}}, \bibinfo {author} {\bibfnamefont
  {S.}~\bibnamefont {Muralithar}}, \bibinfo {author} {\bibfnamefont
  {S.}~\bibnamefont {Nag}}, \bibinfo {author} {\bibfnamefont {S.}~\bibnamefont
  {Nandi}}, \bibinfo {author} {\bibnamefont {Neelam}}, \bibinfo {author}
  {\bibfnamefont {M.~K.}\ \bibnamefont {Raja}}, \bibinfo {author}
  {\bibfnamefont {R.}~\bibnamefont {Raut}}, \bibinfo {author} {\bibfnamefont
  {R.}~\bibnamefont {Santra}}, \bibinfo {author} {\bibfnamefont
  {A.}~\bibnamefont {Sharma}}, \bibinfo {author} {\bibfnamefont
  {S.}~\bibnamefont {Sihotra}}, \bibinfo {author} {\bibfnamefont {A.~K.}\
  \bibnamefont {Singh}}, \bibinfo {author} {\bibfnamefont {R.~P.}\ \bibnamefont
  {Singh}},\ and\ \bibinfo {author} {\bibfnamefont {T.}~\bibnamefont
  {Trivedi}},\ }\bibfield  {title} {\bibinfo {title} {Experimental
  investigation of high-spin states in $^{90}\mathrm{Zr}$},\ }\href
  {https://doi.org/10.1103/PhysRevC.105.044307} {\bibfield  {journal} {\bibinfo
   {journal} {Phys. Rev. C}\ }\textbf {\bibinfo {volume} {105}},\ \bibinfo
  {pages} {044307} (\bibinfo {year} {2022})}\BibitemShut {NoStop}%
\bibitem [{\citenamefont {Chakraborty}\ \emph {et~al.}(2013)\citenamefont
  {Chakraborty}, \citenamefont {Peters}, \citenamefont {Crider}, \citenamefont
  {Andreoiu}, \citenamefont {Bender}, \citenamefont {Cross}, \citenamefont
  {Demand}, \citenamefont {Garnsworthy}, \citenamefont {Garrett}, \citenamefont
  {Hackman}, \citenamefont {Hadinia}, \citenamefont {Ketelhut}, \citenamefont
  {Kumar}, \citenamefont {Leach}, \citenamefont {McEllistrem}, \citenamefont
  {Pore}, \citenamefont {Prados-Est\'evez}, \citenamefont {Rand}, \citenamefont
  {Singh}, \citenamefont {Tardiff}, \citenamefont {Wang}, \citenamefont
  {Wood},\ and\ \citenamefont {Yates}}]{Chakraborty_94Zr}%
  \BibitemOpen
  \bibfield  {author} {\bibinfo {author} {\bibfnamefont {A.}~\bibnamefont
  {Chakraborty}}, \bibinfo {author} {\bibfnamefont {E.~E.}\ \bibnamefont
  {Peters}}, \bibinfo {author} {\bibfnamefont {B.~P.}\ \bibnamefont {Crider}},
  \bibinfo {author} {\bibfnamefont {C.}~\bibnamefont {Andreoiu}}, \bibinfo
  {author} {\bibfnamefont {P.~C.}\ \bibnamefont {Bender}}, \bibinfo {author}
  {\bibfnamefont {D.~S.}\ \bibnamefont {Cross}}, \bibinfo {author}
  {\bibfnamefont {G.~A.}\ \bibnamefont {Demand}}, \bibinfo {author}
  {\bibfnamefont {A.~B.}\ \bibnamefont {Garnsworthy}}, \bibinfo {author}
  {\bibfnamefont {P.~E.}\ \bibnamefont {Garrett}}, \bibinfo {author}
  {\bibfnamefont {G.}~\bibnamefont {Hackman}}, \bibinfo {author} {\bibfnamefont
  {B.}~\bibnamefont {Hadinia}}, \bibinfo {author} {\bibfnamefont
  {S.}~\bibnamefont {Ketelhut}}, \bibinfo {author} {\bibfnamefont
  {A.}~\bibnamefont {Kumar}}, \bibinfo {author} {\bibfnamefont {K.~G.}\
  \bibnamefont {Leach}}, \bibinfo {author} {\bibfnamefont {M.~T.}\ \bibnamefont
  {McEllistrem}}, \bibinfo {author} {\bibfnamefont {J.}~\bibnamefont {Pore}},
  \bibinfo {author} {\bibfnamefont {F.~M.}\ \bibnamefont {Prados-Est\'evez}},
  \bibinfo {author} {\bibfnamefont {E.~T.}\ \bibnamefont {Rand}}, \bibinfo
  {author} {\bibfnamefont {B.}~\bibnamefont {Singh}}, \bibinfo {author}
  {\bibfnamefont {E.~R.}\ \bibnamefont {Tardiff}}, \bibinfo {author}
  {\bibfnamefont {Z.-M.}\ \bibnamefont {Wang}}, \bibinfo {author}
  {\bibfnamefont {J.~L.}\ \bibnamefont {Wood}},\ and\ \bibinfo {author}
  {\bibfnamefont {S.~W.}\ \bibnamefont {Yates}},\ }\bibfield  {title} {\bibinfo
  {title} {Collective structure in $^{94}\mathrm{Zr}$ and subshell effects in
  shape coexistence},\ }\href {https://doi.org/10.1103/PhysRevLett.110.022504}
  {\bibfield  {journal} {\bibinfo  {journal} {Phys. Rev. Lett.}\ }\textbf
  {\bibinfo {volume} {110}},\ \bibinfo {pages} {022504} (\bibinfo {year}
  {2013})}\BibitemShut {NoStop}%
\bibitem [{\citenamefont {Kremer}\ \emph {et~al.}(2016)\citenamefont {Kremer},
  \citenamefont {Aslanidou}, \citenamefont {Bassauer}, \citenamefont {Hilcker},
  \citenamefont {Krugmann}, \citenamefont {von Neumann-Cosel}, \citenamefont
  {Otsuka}, \citenamefont {Pietralla}, \citenamefont {Ponomarev}, \citenamefont
  {Shimizu}, \citenamefont {Singer}, \citenamefont {Steinhilber}, \citenamefont
  {Togashi}, \citenamefont {Tsunoda}, \citenamefont {Werner},\ and\
  \citenamefont {Zweidinger}}]{Kremer_96Zr}%
  \BibitemOpen
  \bibfield  {author} {\bibinfo {author} {\bibfnamefont {C.}~\bibnamefont
  {Kremer}}, \bibinfo {author} {\bibfnamefont {S.}~\bibnamefont {Aslanidou}},
  \bibinfo {author} {\bibfnamefont {S.}~\bibnamefont {Bassauer}}, \bibinfo
  {author} {\bibfnamefont {M.}~\bibnamefont {Hilcker}}, \bibinfo {author}
  {\bibfnamefont {A.}~\bibnamefont {Krugmann}}, \bibinfo {author}
  {\bibfnamefont {P.}~\bibnamefont {von Neumann-Cosel}}, \bibinfo {author}
  {\bibfnamefont {T.}~\bibnamefont {Otsuka}}, \bibinfo {author} {\bibfnamefont
  {N.}~\bibnamefont {Pietralla}}, \bibinfo {author} {\bibfnamefont {V.~Y.}\
  \bibnamefont {Ponomarev}}, \bibinfo {author} {\bibfnamefont {N.}~\bibnamefont
  {Shimizu}}, \bibinfo {author} {\bibfnamefont {M.}~\bibnamefont {Singer}},
  \bibinfo {author} {\bibfnamefont {G.}~\bibnamefont {Steinhilber}}, \bibinfo
  {author} {\bibfnamefont {T.}~\bibnamefont {Togashi}}, \bibinfo {author}
  {\bibfnamefont {Y.}~\bibnamefont {Tsunoda}}, \bibinfo {author} {\bibfnamefont
  {V.}~\bibnamefont {Werner}},\ and\ \bibinfo {author} {\bibfnamefont
  {M.}~\bibnamefont {Zweidinger}},\ }\bibfield  {title} {\bibinfo {title}
  {First measurement of collectivity of coexisting shapes based on {T}ype {II}
  shell evolution: The case of $^{96}\mathrm{Zr}$},\ }\href
  {https://doi.org/10.1103/PhysRevLett.117.172503} {\bibfield  {journal}
  {\bibinfo  {journal} {Phys. Rev. Lett.}\ }\textbf {\bibinfo {volume} {117}},\
  \bibinfo {pages} {172503} (\bibinfo {year} {2016})}\BibitemShut {NoStop}%
\bibitem [{\citenamefont {Garc\'{\i}a-Ramos}\ and\ \citenamefont
  {Heyde}(2020)}]{Garcia-Ramos_Zr}%
  \BibitemOpen
  \bibfield  {author} {\bibinfo {author} {\bibfnamefont {J.~E.}\ \bibnamefont
  {Garc\'{\i}a-Ramos}}\ and\ \bibinfo {author} {\bibfnamefont {K.}~\bibnamefont
  {Heyde}},\ }\bibfield  {title} {\bibinfo {title} {Subtle connection between
  shape coexistence and quantum phase transition: The {Z}r case},\ }\href
  {https://doi.org/10.1103/PhysRevC.102.054333} {\bibfield  {journal} {\bibinfo
   {journal} {Phys. Rev. C}\ }\textbf {\bibinfo {volume} {102}},\ \bibinfo
  {pages} {054333} (\bibinfo {year} {2020})}\BibitemShut {NoStop}%
\bibitem [{\citenamefont {Gavrielov}\ \emph {et~al.}(2022)\citenamefont
  {Gavrielov}, \citenamefont {Leviatan},\ and\ \citenamefont
  {Iachello}}]{Gavrielov_Zr}%
  \BibitemOpen
  \bibfield  {author} {\bibinfo {author} {\bibfnamefont {N.}~\bibnamefont
  {Gavrielov}}, \bibinfo {author} {\bibfnamefont {A.}~\bibnamefont
  {Leviatan}},\ and\ \bibinfo {author} {\bibfnamefont {F.}~\bibnamefont
  {Iachello}},\ }\bibfield  {title} {\bibinfo {title} {Zr isotopes as a region
  of intertwined quantum phase transitions},\ }\href
  {https://doi.org/10.1103/PhysRevC.105.014305} {\bibfield  {journal} {\bibinfo
   {journal} {Phys. Rev. C}\ }\textbf {\bibinfo {volume} {105}},\ \bibinfo
  {pages} {014305} (\bibinfo {year} {2022})}\BibitemShut {NoStop}%
\bibitem [{\citenamefont {Togashi}\ \emph {et~al.}(2016)\citenamefont
  {Togashi}, \citenamefont {Tsunoda}, \citenamefont {Otsuka},\ and\
  \citenamefont {Shimizu}}]{Togashi_Zr}%
  \BibitemOpen
  \bibfield  {author} {\bibinfo {author} {\bibfnamefont {T.}~\bibnamefont
  {Togashi}}, \bibinfo {author} {\bibfnamefont {Y.}~\bibnamefont {Tsunoda}},
  \bibinfo {author} {\bibfnamefont {T.}~\bibnamefont {Otsuka}},\ and\ \bibinfo
  {author} {\bibfnamefont {N.}~\bibnamefont {Shimizu}},\ }\bibfield  {title}
  {\bibinfo {title} {Quantum phase transition in the shape of {Z}r isotopes},\
  }\href {https://doi.org/10.1103/PhysRevLett.117.172502} {\bibfield  {journal}
  {\bibinfo  {journal} {Phys. Rev. Lett.}\ }\textbf {\bibinfo {volume} {117}},\
  \bibinfo {pages} {172502} (\bibinfo {year} {2016})}\BibitemShut {NoStop}%
\bibitem [{\citenamefont {Ball}\ and\ \citenamefont {Fulmer}(1968)}]{90Zr_pd}%
  \BibitemOpen
  \bibfield  {author} {\bibinfo {author} {\bibfnamefont {J.~B.}\ \bibnamefont
  {Ball}}\ and\ \bibinfo {author} {\bibfnamefont {C.~B.}\ \bibnamefont
  {Fulmer}},\ }\bibfield  {title} {\bibinfo {title} {Neutron hole states in
  {Z}=40 nuclei studied with the ($p,d$) reaction on $^{90}${Zr}, $^{91}${Zr},
  and $^{92}${Zr}},\ }\href {https://doi.org/10.1103/PhysRev.172.1199}
  {\bibfield  {journal} {\bibinfo  {journal} {Phys. Rev.}\ }\textbf {\bibinfo
  {volume} {172}},\ \bibinfo {pages} {1199} (\bibinfo {year}
  {1968})}\BibitemShut {NoStop}%
\bibitem [{\citenamefont {Basu}\ and\ \citenamefont
  {McCuthchan}(2020)}]{nds90zr}%
  \BibitemOpen
  \bibfield  {author} {\bibinfo {author} {\bibfnamefont {S.~K.}\ \bibnamefont
  {Basu}}\ and\ \bibinfo {author} {\bibfnamefont {E.~A.}\ \bibnamefont
  {McCuthchan}},\ }\bibfield  {title} {\bibinfo {title} {Nuclear {Data}
  {Sheets} for {A} = 90},\ }\href@noop {} {\bibfield  {journal} {\bibinfo
  {journal} {Nucl. Data Sheets}\ }\textbf {\bibinfo {volume} {165}},\ \bibinfo
  {pages} {1} (\bibinfo {year} {2020})}\BibitemShut {NoStop}%
\bibitem [{\citenamefont {Garrett}\ \emph {et~al.}(2003)\citenamefont
  {Garrett}, \citenamefont {Younes}, \citenamefont {Becker}, \citenamefont
  {Bernstein}, \citenamefont {Baum}, \citenamefont {DiPrete}, \citenamefont
  {Gatenby}, \citenamefont {Johnson}, \citenamefont {McGrath}, \citenamefont
  {Yates}, \citenamefont {Devlin}, \citenamefont {Fotiades}, \citenamefont
  {Nelson},\ and\ \citenamefont {Brown}}]{garrett_90Zr}%
  \BibitemOpen
  \bibfield  {author} {\bibinfo {author} {\bibfnamefont {P.~E.}\ \bibnamefont
  {Garrett}}, \bibinfo {author} {\bibfnamefont {W.}~\bibnamefont {Younes}},
  \bibinfo {author} {\bibfnamefont {J.~A.}\ \bibnamefont {Becker}}, \bibinfo
  {author} {\bibfnamefont {L.~A.}\ \bibnamefont {Bernstein}}, \bibinfo {author}
  {\bibfnamefont {E.~M.}\ \bibnamefont {Baum}}, \bibinfo {author}
  {\bibfnamefont {D.~P.}\ \bibnamefont {DiPrete}}, \bibinfo {author}
  {\bibfnamefont {R.~A.}\ \bibnamefont {Gatenby}}, \bibinfo {author}
  {\bibfnamefont {E.~L.}\ \bibnamefont {Johnson}}, \bibinfo {author}
  {\bibfnamefont {C.~A.}\ \bibnamefont {McGrath}}, \bibinfo {author}
  {\bibfnamefont {S.~W.}\ \bibnamefont {Yates}}, \bibinfo {author}
  {\bibfnamefont {M.}~\bibnamefont {Devlin}}, \bibinfo {author} {\bibfnamefont
  {N.}~\bibnamefont {Fotiades}}, \bibinfo {author} {\bibfnamefont {R.~O.}\
  \bibnamefont {Nelson}},\ and\ \bibinfo {author} {\bibfnamefont {B.~A.}\
  \bibnamefont {Brown}},\ }\bibfield  {title} {\bibinfo {title} {Nuclear
  structure of the closed subshell nucleus ${}^{90}\mathrm{Zr}$ studied with
  the ${(n,n}^{\ensuremath{'}}\ensuremath{\gamma})$ reaction},\ }\href
  {https://doi.org/10.1103/PhysRevC.68.024312} {\bibfield  {journal} {\bibinfo
  {journal} {Phys. Rev. C}\ }\textbf {\bibinfo {volume} {68}},\ \bibinfo
  {pages} {024312} (\bibinfo {year} {2003})}\BibitemShut {NoStop}%
\bibitem [{\citenamefont {Preedom}\ \emph {et~al.}(1968)\citenamefont
  {Preedom}, \citenamefont {Newman},\ and\ \citenamefont {Hiebert}}]{Zr_d3He}%
  \BibitemOpen
  \bibfield  {author} {\bibinfo {author} {\bibfnamefont {B.~M.}\ \bibnamefont
  {Preedom}}, \bibinfo {author} {\bibfnamefont {E.}~\bibnamefont {Newman}},\
  and\ \bibinfo {author} {\bibfnamefont {J.~C.}\ \bibnamefont {Hiebert}},\
  }\bibfield  {title} {\bibinfo {title} {($d$,$^{3}\mathrm{He}$) reactions on
  the even zirconium isotopes},\ }\href
  {https://doi.org/10.1103/PhysRev.166.1156} {\bibfield  {journal} {\bibinfo
  {journal} {Phys. Rev.}\ }\textbf {\bibinfo {volume} {166}},\ \bibinfo {pages}
  {1156} (\bibinfo {year} {1968})}\BibitemShut {NoStop}%
\bibitem [{\citenamefont {Cates}\ \emph {et~al.}(1969)\citenamefont {Cates},
  \citenamefont {Ball},\ and\ \citenamefont {Newman}}]{Zr_3Hed}%
  \BibitemOpen
  \bibfield  {author} {\bibinfo {author} {\bibfnamefont {M.~R.}\ \bibnamefont
  {Cates}}, \bibinfo {author} {\bibfnamefont {J.~B.}\ \bibnamefont {Ball}},\
  and\ \bibinfo {author} {\bibfnamefont {E.}~\bibnamefont {Newman}},\
  }\bibfield  {title} {\bibinfo {title} {Nuclear-structure studies of
  $^{91,92,93,95,97}\mathrm{Nb}$ and $^{94}\mathrm{Mo}$ with the
  ($^{3}\mathrm{He}$, $d$) reaction},\ }\href
  {https://doi.org/10.1103/PhysRev.187.1682} {\bibfield  {journal} {\bibinfo
  {journal} {Phys. Rev.}\ }\textbf {\bibinfo {volume} {187}},\ \bibinfo {pages}
  {1682} (\bibinfo {year} {1969})}\BibitemShut {NoStop}%
\bibitem [{\citenamefont {Saha}\ \emph {et~al.}(1979)\citenamefont {Saha},
  \citenamefont {Jones}, \citenamefont {Put},\ and\ \citenamefont
  {Siemssen}}]{Saha}%
  \BibitemOpen
  \bibfield  {author} {\bibinfo {author} {\bibfnamefont {A.}~\bibnamefont
  {Saha}}, \bibinfo {author} {\bibfnamefont {G.~D.}\ \bibnamefont {Jones}},
  \bibinfo {author} {\bibfnamefont {L.~W.}\ \bibnamefont {Put}},\ and\ \bibinfo
  {author} {\bibfnamefont {R.~H.}\ \bibnamefont {Siemssen}},\ }\bibfield
  {title} {\bibinfo {title} {Unusually strong excitation of the 1.59 {M}e{V}
  $0^+$ state in the $^{100}${M}o($d$,$^6${L}i)$^{96}${Z}r reaction},\ }\href
  {https://doi.org/https://doi.org/10.1016/0370-2693(79)90737-8} {\bibfield
  {journal} {\bibinfo  {journal} {Phys. Lett. B}\ }\textbf {\bibinfo {volume}
  {82}},\ \bibinfo {pages} {208} (\bibinfo {year} {1979})}\BibitemShut
  {NoStop}%
\bibitem [{\citenamefont {Sieja}\ \emph {et~al.}(2009)\citenamefont {Sieja},
  \citenamefont {Nowacki}, \citenamefont {Langanke},\ and\ \citenamefont
  {Mart\'{\i}nez-Pinedo}}]{Sieja_Zr_SM}%
  \BibitemOpen
  \bibfield  {author} {\bibinfo {author} {\bibfnamefont {K.}~\bibnamefont
  {Sieja}}, \bibinfo {author} {\bibfnamefont {F.}~\bibnamefont {Nowacki}},
  \bibinfo {author} {\bibfnamefont {K.}~\bibnamefont {Langanke}},\ and\
  \bibinfo {author} {\bibfnamefont {G.}~\bibnamefont {Mart\'{\i}nez-Pinedo}},\
  }\bibfield  {title} {\bibinfo {title} {Shell model description of zirconium
  isotopes},\ }\href {https://doi.org/10.1103/PhysRevC.79.064310} {\bibfield
  {journal} {\bibinfo  {journal} {Phys. Rev. C}\ }\textbf {\bibinfo {volume}
  {79}},\ \bibinfo {pages} {064310} (\bibinfo {year} {2009})}\BibitemShut
  {NoStop}%
\bibitem [{\citenamefont {Mach}\ \emph {et~al.}(1990)\citenamefont {Mach},
  \citenamefont {Warburton}, \citenamefont {Krips}, \citenamefont {Gill},\ and\
  \citenamefont {Moszy\ifmmode~\acute{n}\else \'{n}\fi{}ski}}]{Mach_Zr}%
  \BibitemOpen
  \bibfield  {author} {\bibinfo {author} {\bibfnamefont {H.}~\bibnamefont
  {Mach}}, \bibinfo {author} {\bibfnamefont {E.~K.}\ \bibnamefont {Warburton}},
  \bibinfo {author} {\bibfnamefont {W.}~\bibnamefont {Krips}}, \bibinfo
  {author} {\bibfnamefont {R.~L.}\ \bibnamefont {Gill}},\ and\ \bibinfo
  {author} {\bibfnamefont {M.}~\bibnamefont {Moszy\ifmmode~\acute{n}\else
  \'{n}\fi{}ski}},\ }\bibfield  {title} {\bibinfo {title} {Evidence for the
  contribution of 0${\mathit{f}}_{5/2}$,1${\mathit{p}}_{3/2}$ proton
  excitations in the low-lying states in $^{92,94}\mathrm{Zr}$},\ }\href
  {https://doi.org/10.1103/PhysRevC.42.568} {\bibfield  {journal} {\bibinfo
  {journal} {Phys. Rev. C}\ }\textbf {\bibinfo {volume} {42}},\ \bibinfo
  {pages} {568} (\bibinfo {year} {1990})}\BibitemShut {NoStop}%
\bibitem [{\citenamefont {Marchini}\ \emph {et~al.}(2026)\citenamefont
  {Marchini}, \citenamefont {Rocchini}, \citenamefont {Zielińska},
  \citenamefont {Nannini}, \citenamefont {Doherty}, \citenamefont {Gavrielov},
  \citenamefont {Garrett}, \citenamefont {Hady{\'n}ska-Kl{\c e}k},
  \citenamefont {Goasduff}, \citenamefont {Testov}, \citenamefont {Bakes},
  \citenamefont {Bazzacco}, \citenamefont {Benzoni}, \citenamefont {Berry},
  \citenamefont {Brugnara}, \citenamefont {Camera}, \citenamefont {Catford},
  \citenamefont {Chiari}, \citenamefont {Galtarossa}, \citenamefont {Gelli},
  \citenamefont {Gottardo}, \citenamefont {Gozzelino}, \citenamefont {Illana},
  \citenamefont {Keatings}, \citenamefont {Mengoni}, \citenamefont {Morrison},
  \citenamefont {Napoli}, \citenamefont {Ottanelli}, \citenamefont {Ottanelli},
  \citenamefont {Pasqualato}, \citenamefont {Recchia}, \citenamefont
  {Riccetto}, \citenamefont {Scheck}, \citenamefont {Siciliano}, \citenamefont
  {Valiente~D\'obon},\ and\ \citenamefont {Zanon}}]{Naomi_94Zr}%
  \BibitemOpen
  \bibfield  {author} {\bibinfo {author} {\bibfnamefont {N.}~\bibnamefont
  {Marchini}}, \bibinfo {author} {\bibfnamefont {M.}~\bibnamefont {Rocchini}},
  \bibinfo {author} {\bibfnamefont {M.}~\bibnamefont {Zielińska}}, \bibinfo
  {author} {\bibfnamefont {A.}~\bibnamefont {Nannini}}, \bibinfo {author}
  {\bibfnamefont {D.~T.}\ \bibnamefont {Doherty}}, \bibinfo {author}
  {\bibfnamefont {N.}~\bibnamefont {Gavrielov}}, \bibinfo {author}
  {\bibfnamefont {P.~E.}\ \bibnamefont {Garrett}}, \bibinfo {author}
  {\bibfnamefont {K.}~\bibnamefont {Hady{\'n}ska-Kl{\c e}k}}, \bibinfo {author}
  {\bibfnamefont {A.}~\bibnamefont {Goasduff}}, \bibinfo {author}
  {\bibfnamefont {D.}~\bibnamefont {Testov}}, \bibinfo {author} {\bibfnamefont
  {S.~D.}\ \bibnamefont {Bakes}}, \bibinfo {author} {\bibfnamefont
  {D.}~\bibnamefont {Bazzacco}}, \bibinfo {author} {\bibfnamefont
  {G.}~\bibnamefont {Benzoni}}, \bibinfo {author} {\bibfnamefont
  {T.}~\bibnamefont {Berry}}, \bibinfo {author} {\bibfnamefont
  {D.}~\bibnamefont {Brugnara}}, \bibinfo {author} {\bibfnamefont
  {F.}~\bibnamefont {Camera}}, \bibinfo {author} {\bibfnamefont {W.~N.}\
  \bibnamefont {Catford}}, \bibinfo {author} {\bibfnamefont {M.}~\bibnamefont
  {Chiari}}, \bibinfo {author} {\bibfnamefont {F.}~\bibnamefont {Galtarossa}},
  \bibinfo {author} {\bibfnamefont {N.}~\bibnamefont {Gelli}}, \bibinfo
  {author} {\bibfnamefont {A.}~\bibnamefont {Gottardo}}, \bibinfo {author}
  {\bibfnamefont {A.}~\bibnamefont {Gozzelino}}, \bibinfo {author}
  {\bibfnamefont {A.}~\bibnamefont {Illana}}, \bibinfo {author} {\bibfnamefont
  {J.}~\bibnamefont {Keatings}}, \bibinfo {author} {\bibfnamefont
  {D.}~\bibnamefont {Mengoni}}, \bibinfo {author} {\bibfnamefont
  {L.}~\bibnamefont {Morrison}}, \bibinfo {author} {\bibfnamefont {D.~R.}\
  \bibnamefont {Napoli}}, \bibinfo {author} {\bibfnamefont {M.}~\bibnamefont
  {Ottanelli}}, \bibinfo {author} {\bibfnamefont {P.}~\bibnamefont
  {Ottanelli}}, \bibinfo {author} {\bibfnamefont {G.}~\bibnamefont
  {Pasqualato}}, \bibinfo {author} {\bibfnamefont {F.}~\bibnamefont {Recchia}},
  \bibinfo {author} {\bibfnamefont {S.}~\bibnamefont {Riccetto}}, \bibinfo
  {author} {\bibfnamefont {M.}~\bibnamefont {Scheck}}, \bibinfo {author}
  {\bibfnamefont {M.}~\bibnamefont {Siciliano}}, \bibinfo {author}
  {\bibfnamefont {J.~J.}\ \bibnamefont {Valiente~D\'obon}},\ and\ \bibinfo
  {author} {\bibfnamefont {I.}~\bibnamefont {Zanon}},\ }\bibfield  {title}
  {\bibinfo {title} {Shape coexistence in $^{94}\mathrm{Zr}$ from a
  model-independent analysis},\ }\href
  {https://doi.org/10.1016/j.physletb.2026.140668} {\bibfield  {journal}
  {\bibinfo  {journal} {Phys. Lett. B}\ }\textbf {\bibinfo {volume} {879}},\
  \bibinfo {pages} {140668} (\bibinfo {year} {2026})}\BibitemShut {NoStop}%
\bibitem [{\citenamefont {Zieli{\'n}ska}\ \emph {et~al.}(2026)\citenamefont
  {Zieli{\'n}ska}, \citenamefont {Garrett}, \citenamefont {Rocchini},
  \citenamefont {Pannu}, \citenamefont {Bergmaier}, \citenamefont {Bildstein},
  \citenamefont {Calder}, \citenamefont {Dworatzek}, \citenamefont {Vedia},
  \citenamefont {Ahmed}, \citenamefont {Andreoiu}, \citenamefont {Annen},
  \citenamefont {Asch}, \citenamefont {Buck}, \citenamefont {Coleman},
  \citenamefont {{Diaz Varela}}, \citenamefont {Doherty}, \citenamefont {Avaa},
  \citenamefont {Ball}, \citenamefont {Dowie}, \citenamefont {Faestermann},
  \citenamefont {Garnsworthy}, \citenamefont {Geerlof}, \citenamefont
  {Greaves}, \citenamefont {Grimes}, \citenamefont {Hackman}, \citenamefont
  {Hertenberger}, \citenamefont {Hicks}, \citenamefont {Kalaydjieva},
  \citenamefont {Karayonchev}, \citenamefont {{La Marca}}, \citenamefont
  {Lange}, \citenamefont {Lenardo}, \citenamefont {Maqungo}, \citenamefont
  {Marchini}, \citenamefont {Marlow}, \citenamefont {Martin}, \citenamefont
  {Mashtakov}, \citenamefont {McCourt}, \citenamefont {Murias}, \citenamefont
  {Murillo-Morales}, \citenamefont {Nannini}, \citenamefont {Peters},
  \citenamefont {Petrache}, \citenamefont {Porzio}, \citenamefont {Richardson},
  \citenamefont {Scheck}, \citenamefont {Spagnoletti}, \citenamefont
  {Stoychev}, \citenamefont {Svensson}, \citenamefont {Taddei}, \citenamefont
  {Tocabens}, \citenamefont {Triambak}, \citenamefont {Vandebrouck},
  \citenamefont {Umashankar}, \citenamefont {Valbuena}, \citenamefont
  {Wrzosek-Lipska}, \citenamefont {Wu},\ and\ \citenamefont
  {Zidar}}]{Zielinska_96Zr}%
  \BibitemOpen
  \bibfield  {author} {\bibinfo {author} {\bibfnamefont {M.}~\bibnamefont
  {Zieli{\'n}ska}}, \bibinfo {author} {\bibfnamefont {P.~E.}\ \bibnamefont
  {Garrett}}, \bibinfo {author} {\bibfnamefont {M.}~\bibnamefont {Rocchini}},
  \bibinfo {author} {\bibfnamefont {S.}~\bibnamefont {Pannu}}, \bibinfo
  {author} {\bibfnamefont {A.}~\bibnamefont {Bergmaier}}, \bibinfo {author}
  {\bibfnamefont {V.}~\bibnamefont {Bildstein}}, \bibinfo {author}
  {\bibfnamefont {A.~B.}\ \bibnamefont {Calder}}, \bibinfo {author}
  {\bibfnamefont {S.~R.}\ \bibnamefont {Dworatzek}}, \bibinfo {author}
  {\bibfnamefont {V.}~\bibnamefont {Vedia}}, \bibinfo {author} {\bibfnamefont
  {Z.~T.}\ \bibnamefont {Ahmed}}, \bibinfo {author} {\bibfnamefont
  {C.}~\bibnamefont {Andreoiu}}, \bibinfo {author} {\bibfnamefont {D.~W.}\
  \bibnamefont {Annen}}, \bibinfo {author} {\bibfnamefont {H.}~\bibnamefont
  {Asch}}, \bibinfo {author} {\bibfnamefont {S.}~\bibnamefont {Buck}}, \bibinfo
  {author} {\bibfnamefont {R.~J.}\ \bibnamefont {Coleman}}, \bibinfo {author}
  {\bibfnamefont {A.}~\bibnamefont {{Diaz Varela}}}, \bibinfo {author}
  {\bibfnamefont {D.~T.}\ \bibnamefont {Doherty}}, \bibinfo {author}
  {\bibfnamefont {A.~A.}\ \bibnamefont {Avaa}}, \bibinfo {author}
  {\bibfnamefont {G.~C.}\ \bibnamefont {Ball}}, \bibinfo {author}
  {\bibfnamefont {J.~T.~H.}\ \bibnamefont {Dowie}}, \bibinfo {author}
  {\bibfnamefont {T.}~\bibnamefont {Faestermann}}, \bibinfo {author}
  {\bibfnamefont {A.~B.}\ \bibnamefont {Garnsworthy}}, \bibinfo {author}
  {\bibfnamefont {E.~D.}\ \bibnamefont {Geerlof}}, \bibinfo {author}
  {\bibfnamefont {B.}~\bibnamefont {Greaves}}, \bibinfo {author} {\bibfnamefont
  {A.~L.}\ \bibnamefont {Grimes}}, \bibinfo {author} {\bibfnamefont
  {G.}~\bibnamefont {Hackman}}, \bibinfo {author} {\bibfnamefont
  {R.}~\bibnamefont {Hertenberger}}, \bibinfo {author} {\bibfnamefont {S.~F.}\
  \bibnamefont {Hicks}}, \bibinfo {author} {\bibfnamefont {D.}~\bibnamefont
  {Kalaydjieva}}, \bibinfo {author} {\bibfnamefont {V.}~\bibnamefont
  {Karayonchev}}, \bibinfo {author} {\bibfnamefont {T.}~\bibnamefont {{La
  Marca}}}, \bibinfo {author} {\bibfnamefont {S.}~\bibnamefont {Lange}},
  \bibinfo {author} {\bibfnamefont {B.}~\bibnamefont {Lenardo}}, \bibinfo
  {author} {\bibfnamefont {L.}~\bibnamefont {Maqungo}}, \bibinfo {author}
  {\bibfnamefont {N.}~\bibnamefont {Marchini}}, \bibinfo {author}
  {\bibfnamefont {B.}~\bibnamefont {Marlow}}, \bibinfo {author} {\bibfnamefont
  {M.~S.}\ \bibnamefont {Martin}}, \bibinfo {author} {\bibfnamefont {K.~R.}\
  \bibnamefont {Mashtakov}}, \bibinfo {author} {\bibfnamefont {E.}~\bibnamefont
  {McCourt}}, \bibinfo {author} {\bibfnamefont {J.~R.}\ \bibnamefont {Murias}},
  \bibinfo {author} {\bibfnamefont {S.}~\bibnamefont {Murillo-Morales}},
  \bibinfo {author} {\bibfnamefont {A.}~\bibnamefont {Nannini}}, \bibinfo
  {author} {\bibfnamefont {E.~E.}\ \bibnamefont {Peters}}, \bibinfo {author}
  {\bibfnamefont {C.~M.}\ \bibnamefont {Petrache}}, \bibinfo {author}
  {\bibfnamefont {C.}~\bibnamefont {Porzio}}, \bibinfo {author} {\bibfnamefont
  {G.}~\bibnamefont {Richardson}}, \bibinfo {author} {\bibfnamefont
  {M.}~\bibnamefont {Scheck}}, \bibinfo {author} {\bibfnamefont
  {P.}~\bibnamefont {Spagnoletti}}, \bibinfo {author} {\bibfnamefont
  {K.}~\bibnamefont {Stoychev}}, \bibinfo {author} {\bibfnamefont {C.~E.}\
  \bibnamefont {Svensson}}, \bibinfo {author} {\bibfnamefont {E.}~\bibnamefont
  {Taddei}}, \bibinfo {author} {\bibfnamefont {G.}~\bibnamefont {Tocabens}},
  \bibinfo {author} {\bibfnamefont {S.}~\bibnamefont {Triambak}}, \bibinfo
  {author} {\bibfnamefont {M.}~\bibnamefont {Vandebrouck}}, \bibinfo {author}
  {\bibfnamefont {R.}~\bibnamefont {Umashankar}}, \bibinfo {author}
  {\bibfnamefont {S.}~\bibnamefont {Valbuena}}, \bibinfo {author}
  {\bibfnamefont {K.}~\bibnamefont {Wrzosek-Lipska}}, \bibinfo {author}
  {\bibfnamefont {F.}~\bibnamefont {Wu}},\ and\ \bibinfo {author}
  {\bibfnamefont {T.}~\bibnamefont {Zidar}},\ }\bibfield  {title} {\bibinfo
  {title} {Collectivity of the deformed structure in $^{96}${Z}r from
  complementary {C}oulomb-excitation and $\beta$-decay measurements},\ }\href
  {https://doi.org/https://doi.org/10.1016/j.physletb.2026.140481} {\bibfield
  {journal} {\bibinfo  {journal} {Phys. Lett. B}\ }\textbf {\bibinfo {volume}
  {877}},\ \bibinfo {pages} {140481} (\bibinfo {year} {2026})}\BibitemShut
  {NoStop}%
\bibitem [{\citenamefont {Petrovici}\ and\ \citenamefont
  {Mare}(2020)}]{Petrovici_96Zr}%
  \BibitemOpen
  \bibfield  {author} {\bibinfo {author} {\bibfnamefont {A.}~\bibnamefont
  {Petrovici}}\ and\ \bibinfo {author} {\bibfnamefont {A.~S.}\ \bibnamefont
  {Mare}},\ }\bibfield  {title} {\bibinfo {title} {Triple shape coexistence and
  $\ensuremath{\beta}$ decay of $^{96}\mathrm{Y}$ to $^{96}\mathrm{Zr}$},\
  }\href {https://doi.org/10.1103/PhysRevC.101.024307} {\bibfield  {journal}
  {\bibinfo  {journal} {Phys. Rev. C}\ }\textbf {\bibinfo {volume} {101}},\
  \bibinfo {pages} {024307} (\bibinfo {year} {2020})}\BibitemShut {NoStop}%
\bibitem [{\citenamefont {Sheline}\ \emph {et~al.}(1972)\citenamefont
  {Sheline}, \citenamefont {Ragnarsson},\ and\ \citenamefont
  {Nilsson}}]{Sheline_1972}%
  \BibitemOpen
  \bibfield  {author} {\bibinfo {author} {\bibfnamefont {R.~K.}\ \bibnamefont
  {Sheline}}, \bibinfo {author} {\bibfnamefont {I.}~\bibnamefont
  {Ragnarsson}},\ and\ \bibinfo {author} {\bibfnamefont {S.~G.}\ \bibnamefont
  {Nilsson}},\ }\bibfield  {title} {\bibinfo {title} {Shell structure for
  deformed nuclear shapes},\ }\href
  {https://doi.org/10.1016/0370-2693(72)90440-6} {\bibfield  {journal}
  {\bibinfo  {journal} {Phys. Lett. B}\ }\textbf {\bibinfo {volume} {41}},\
  \bibinfo {pages} {115} (\bibinfo {year} {1972})}\BibitemShut {NoStop}%
\bibitem [{\citenamefont {Sugawara}\ \emph {et~al.}(2017)\citenamefont
  {Sugawara}, \citenamefont {Toh}, \citenamefont {Koizumi}, \citenamefont
  {Oshima}, \citenamefont {Kimura}, \citenamefont {Kin}, \citenamefont
  {Hatsukawa},\ and\ \citenamefont {Kusakari}}]{Sugawara_92Zr}%
  \BibitemOpen
  \bibfield  {author} {\bibinfo {author} {\bibfnamefont {M.}~\bibnamefont
  {Sugawara}}, \bibinfo {author} {\bibfnamefont {Y.}~\bibnamefont {Toh}},
  \bibinfo {author} {\bibfnamefont {M.}~\bibnamefont {Koizumi}}, \bibinfo
  {author} {\bibfnamefont {M.}~\bibnamefont {Oshima}}, \bibinfo {author}
  {\bibfnamefont {A.}~\bibnamefont {Kimura}}, \bibinfo {author} {\bibfnamefont
  {T.}~\bibnamefont {Kin}}, \bibinfo {author} {\bibfnamefont {Y.}~\bibnamefont
  {Hatsukawa}},\ and\ \bibinfo {author} {\bibfnamefont {H.}~\bibnamefont
  {Kusakari}},\ }\bibfield  {title} {\bibinfo {title} {Intermediate-spin states
  of ${}^{92}${Zr} and a large {B(E2)} value between the ${10}_{1}^{+}$ and
  ${8}_{1}^{+}$ states},\ }\href {https://doi.org/10.1103/PhysRevC.96.024314}
  {\bibfield  {journal} {\bibinfo  {journal} {Phys. Rev. C}\ }\textbf {\bibinfo
  {volume} {96}},\ \bibinfo {pages} {024314} (\bibinfo {year}
  {2017})}\BibitemShut {NoStop}%
\bibitem [{\citenamefont {Buckley}\ \emph {et~al.}(2000)\citenamefont
  {Buckley}, \citenamefont {Huser}, \citenamefont {Jivan}, \citenamefont
  {Chun},\ and\ \citenamefont {Ruth}}]{tr13}%
  \BibitemOpen
  \bibfield  {author} {\bibinfo {author} {\bibfnamefont {K.~R.}\ \bibnamefont
  {Buckley}}, \bibinfo {author} {\bibfnamefont {J.~M.}\ \bibnamefont {Huser}},
  \bibinfo {author} {\bibfnamefont {S.}~\bibnamefont {Jivan}}, \bibinfo
  {author} {\bibfnamefont {K.~S.}\ \bibnamefont {Chun}},\ and\ \bibinfo
  {author} {\bibfnamefont {T.~J.}\ \bibnamefont {Ruth}},\ }\bibfield  {title}
  {\bibinfo {title} {$^{11}${C}-methane production in small volume, high
  pressure gas targets},\ }\href
  {https://doi.org/doi:10.1524/ract.2000.88.3-4.201} {\bibfield  {journal}
  {\bibinfo  {journal} {Radiochimica Acta}\ }\textbf {\bibinfo {volume} {88}},\
  \bibinfo {pages} {201} (\bibinfo {year} {2000})}\BibitemShut {NoStop}%
\bibitem [{\citenamefont {Zeisler}\ \emph {et~al.}(2019)\citenamefont
  {Zeisler}, \citenamefont {Clarke}, \citenamefont {Kumlin}, \citenamefont
  {Hook}, \citenamefont {Varah},\ and\ \citenamefont {Höhr}}]{tr13_targetsys}%
  \BibitemOpen
  \bibfield  {author} {\bibinfo {author} {\bibfnamefont {S.}~\bibnamefont
  {Zeisler}}, \bibinfo {author} {\bibfnamefont {B.}~\bibnamefont {Clarke}},
  \bibinfo {author} {\bibfnamefont {J.}~\bibnamefont {Kumlin}}, \bibinfo
  {author} {\bibfnamefont {B.}~\bibnamefont {Hook}}, \bibinfo {author}
  {\bibfnamefont {S.}~\bibnamefont {Varah}},\ and\ \bibinfo {author}
  {\bibfnamefont {C.}~\bibnamefont {Höhr}},\ }\bibfield  {title} {\bibinfo
  {title} {A {Compact} {Quick}-{Release} {Solid} {Target} {System} for the
  {TRIUMF} {TR13} {Cyclotron}},\ }\href
  {https://doi.org/10.3390/instruments3010016} {\bibfield  {journal} {\bibinfo
  {journal} {Instruments}\ }\textbf {\bibinfo {volume} {3}},\ \bibinfo {pages}
  {16} (\bibinfo {year} {2019})}\BibitemShut {NoStop}%
\bibitem [{\citenamefont {Garnsworthy}\ \emph {et~al.}(2019)\citenamefont
  {Garnsworthy} \emph {et~al.}}]{griffin2019}%
  \BibitemOpen
  \bibfield  {author} {\bibinfo {author} {\bibfnamefont {A.~B.}\ \bibnamefont
  {Garnsworthy}} \emph {et~al.},\ }\bibfield  {title} {\bibinfo {title} {The
  {GRIFFIN} facility for {D}ecay-spectroscopy studies at {TRIUMF-ISAC}},\
  }\href@noop {} {\bibfield  {journal} {\bibinfo  {journal} {Nuclear
  Instruments and Methods in Physics Research Section A}\ }\textbf {\bibinfo
  {volume} {918}},\ \bibinfo {pages} {9} (\bibinfo {year} {2019})}\BibitemShut
  {NoStop}%
\bibitem [{\citenamefont {Kib\'{e}di}\ \emph {et~al.}(2008)\citenamefont
  {Kib\'{e}di}, \citenamefont {Burrows}, \citenamefont {Trzhaskovskaya},
  \citenamefont {Davidson},\ and\ \citenamefont {C.W.~Nestor}}]{2008Ki07}%
  \BibitemOpen
  \bibfield  {author} {\bibinfo {author} {\bibfnamefont {T.}~\bibnamefont
  {Kib\'{e}di}}, \bibinfo {author} {\bibfnamefont {T.}~\bibnamefont {Burrows}},
  \bibinfo {author} {\bibfnamefont {M.}~\bibnamefont {Trzhaskovskaya}},
  \bibinfo {author} {\bibfnamefont {P.}~\bibnamefont {Davidson}},\ and\
  \bibinfo {author} {\bibfnamefont {J.}~\bibnamefont {C.W.~Nestor}},\
  }\bibfield  {title} {\bibinfo {title} {Evaluation of theoretical conversion
  coefficients using {BrIcc}},\ }\href@noop {} {\bibfield  {journal} {\bibinfo
  {journal} {Nuclear Instruments and Methods in Physics Research Section A}\
  }\textbf {\bibinfo {volume} {589}},\ \bibinfo {pages} {202} (\bibinfo {year}
  {2008})}\BibitemShut {NoStop}%
\bibitem [{\citenamefont {Baglin}(2012)}]{nds92}%
  \BibitemOpen
  \bibfield  {author} {\bibinfo {author} {\bibfnamefont {C.~M.}\ \bibnamefont
  {Baglin}},\ }\bibfield  {title} {\bibinfo {title} {Nuclear {Data} {Sheets}
  for {A} = 92},\ }\href {https://doi.org/10.1016/j.nds.2012.10.001} {\bibfield
   {journal} {\bibinfo  {journal} {Nuclear Data Sheets}\ }\textbf {\bibinfo
  {volume} {113}},\ \bibinfo {pages} {2187} (\bibinfo {year}
  {2012})}\BibitemShut {NoStop}%
\bibitem [{\citenamefont {Bhatia}\ \emph {et~al.}(1971)\citenamefont {Bhatia},
  \citenamefont {Daehnick},\ and\ \citenamefont {Canada}}]{bhatia1971}%
  \BibitemOpen
  \bibfield  {author} {\bibinfo {author} {\bibfnamefont {T.~S.}\ \bibnamefont
  {Bhatia}}, \bibinfo {author} {\bibfnamefont {W.~W.}\ \bibnamefont
  {Daehnick}},\ and\ \bibinfo {author} {\bibfnamefont {T.~R.}\ \bibnamefont
  {Canada}},\ }\bibfield  {title} {\bibinfo {title} {High-{Resolution} {Study}
  of {$^{92}$}{N}b by (d,t) and (d,t$\gamma$) {Reactions}},\ }\href
  {https://doi.org/10.1103/PhysRevC.3.1361} {\bibfield  {journal} {\bibinfo
  {journal} {Phys. Rev. C}\ }\textbf {\bibinfo {volume} {3}},\ \bibinfo {pages}
  {1361} (\bibinfo {year} {1971})}\BibitemShut {NoStop}%
\bibitem [{\citenamefont {Rumore}\ \emph {et~al.}(1984)\citenamefont {Rumore},
  \citenamefont {Dickey}, \citenamefont {Fields},\ and\ \citenamefont
  {Kraushaar}}]{rumore1984}%
  \BibitemOpen
  \bibfield  {author} {\bibinfo {author} {\bibfnamefont {M.~A.}\ \bibnamefont
  {Rumore}}, \bibinfo {author} {\bibfnamefont {S.~A.}\ \bibnamefont {Dickey}},
  \bibinfo {author} {\bibfnamefont {C.~A.}\ \bibnamefont {Fields}},\ and\
  \bibinfo {author} {\bibfnamefont {J.~J.}\ \bibnamefont {Kraushaar}},\
  }\bibfield  {title} {\bibinfo {title} {The {$^{93}$Nb}(p, d){$^{92}$Nb}
  reaction at 26.3 {MeV}},\ }\href
  {https://doi.org/10.1016/0375-9474(84)90594-3} {\bibfield  {journal}
  {\bibinfo  {journal} {Nucl. Phys. A}\ }\textbf {\bibinfo {volume} {423}},\
  \bibinfo {pages} {350} (\bibinfo {year} {1984})}\BibitemShut {NoStop}%
\bibitem [{\citenamefont {Smith}\ \emph {et~al.}(2019)\citenamefont {Smith},
  \citenamefont {MacLean}, \citenamefont {Ashfield}, \citenamefont {Chester},
  \citenamefont {Garnsworthy},\ and\ \citenamefont
  {Svensson}}]{smith_gamma-gamma_2019}%
  \BibitemOpen
  \bibfield  {author} {\bibinfo {author} {\bibfnamefont {J.~K.}\ \bibnamefont
  {Smith}}, \bibinfo {author} {\bibfnamefont {A.~D.}\ \bibnamefont {MacLean}},
  \bibinfo {author} {\bibfnamefont {W.}~\bibnamefont {Ashfield}}, \bibinfo
  {author} {\bibfnamefont {A.}~\bibnamefont {Chester}}, \bibinfo {author}
  {\bibfnamefont {A.~B.}\ \bibnamefont {Garnsworthy}},\ and\ \bibinfo {author}
  {\bibfnamefont {C.~E.}\ \bibnamefont {Svensson}},\ }\bibfield  {title}
  {\bibinfo {title} {Gamma-gamma angular correlation analysis techniques with
  the {GRIFFIN} spectrometer},\ }\href
  {https://doi.org/https://doi.org/10.1016/j.nima.2018.10.097} {\bibfield
  {journal} {\bibinfo  {journal} {Nuclear Instruments and Methods in Physics
  Research Section A: Accelerators, Spectrometers, Detectors and Associated
  Equipment}\ }\textbf {\bibinfo {volume} {922}},\ \bibinfo {pages} {47}
  (\bibinfo {year} {2019})}\BibitemShut {NoStop}%
\bibitem [{\citenamefont {Krane}\ and\ \citenamefont
  {Steffen}(1970)}]{krane_determination_1970}%
  \BibitemOpen
  \bibfield  {author} {\bibinfo {author} {\bibfnamefont {K.~S.}\ \bibnamefont
  {Krane}}\ and\ \bibinfo {author} {\bibfnamefont {R.~M.}\ \bibnamefont
  {Steffen}},\ }\bibfield  {title} {\bibinfo {title} {Determination of the
  {E2}/{M1} {Multipole} {Mixing} {Ratios} of the {Gamma} {Transitions} in
  {Cd$^{110}$}},\ }\href {https://doi.org/10.1103/PhysRevC.2.724} {\bibfield
  {journal} {\bibinfo  {journal} {Phys. Rev. C}\ }\textbf {\bibinfo {volume}
  {2}},\ \bibinfo {pages} {724} (\bibinfo {year} {1970})}\BibitemShut {NoStop}%
\bibitem [{\citenamefont {Bunker}\ \emph {et~al.}(1962)\citenamefont {Bunker},
  \citenamefont {Dropesky}, \citenamefont {Knight},\ and\ \citenamefont
  {Starner}}]{bunker_decay_1962}%
  \BibitemOpen
  \bibfield  {author} {\bibinfo {author} {\bibfnamefont {M.~E.}\ \bibnamefont
  {Bunker}}, \bibinfo {author} {\bibfnamefont {B.~J.}\ \bibnamefont
  {Dropesky}}, \bibinfo {author} {\bibfnamefont {J.~D.}\ \bibnamefont
  {Knight}},\ and\ \bibinfo {author} {\bibfnamefont {J.~W.}\ \bibnamefont
  {Starner}},\ }\bibfield  {title} {\bibinfo {title} {Decay of {Y$^{92}$} and
  {Nb$^{92}$}},\ }\href {https://doi.org/10.1103/PhysRev.127.844} {\bibfield
  {journal} {\bibinfo  {journal} {Phys. Rev.}\ }\textbf {\bibinfo {volume}
  {127}},\ \bibinfo {pages} {844} (\bibinfo {year} {1962})}\BibitemShut
  {NoStop}%
\bibitem [{\citenamefont {Fransen}\ \emph {et~al.}(2005)\citenamefont
  {Fransen}, \citenamefont {Werner}, \citenamefont {Bandyopadhyay},
  \citenamefont {Boukharouba}, \citenamefont {Lesher}, \citenamefont
  {McEllistrem}, \citenamefont {Jolie}, \citenamefont {Pietralla},
  \citenamefont {Brentano},\ and\ \citenamefont {Yates}}]{Fransen_92Zr}%
  \BibitemOpen
  \bibfield  {author} {\bibinfo {author} {\bibfnamefont {C.}~\bibnamefont
  {Fransen}}, \bibinfo {author} {\bibfnamefont {V.}~\bibnamefont {Werner}},
  \bibinfo {author} {\bibfnamefont {D.}~\bibnamefont {Bandyopadhyay}}, \bibinfo
  {author} {\bibfnamefont {N.}~\bibnamefont {Boukharouba}}, \bibinfo {author}
  {\bibfnamefont {S.~R.}\ \bibnamefont {Lesher}}, \bibinfo {author}
  {\bibfnamefont {M.~T.}\ \bibnamefont {McEllistrem}}, \bibinfo {author}
  {\bibfnamefont {J.}~\bibnamefont {Jolie}}, \bibinfo {author} {\bibfnamefont
  {N.}~\bibnamefont {Pietralla}}, \bibinfo {author} {\bibfnamefont {P.~v.}\
  \bibnamefont {Brentano}},\ and\ \bibinfo {author} {\bibfnamefont {S.~W.}\
  \bibnamefont {Yates}},\ }\bibfield  {title} {\bibinfo {title} {Investigation
  of low-spin states in $^{92}\mathrm{Zr}$ with the
  ($n,{n}^{'}\ensuremath{\gamma}$) reaction},\ }\href
  {https://doi.org/10.1103/PhysRevC.71.054304} {\bibfield  {journal} {\bibinfo
  {journal} {Phys. Rev. C}\ }\textbf {\bibinfo {volume} {71}},\ \bibinfo
  {pages} {054304} (\bibinfo {year} {2005})}\BibitemShut {NoStop}%
\bibitem [{\citenamefont {Peters}\ \emph {et~al.}(2013)\citenamefont {Peters},
  \citenamefont {Chakraborty}, \citenamefont {Crider}, \citenamefont {Davis},
  \citenamefont {Gnanamani}, \citenamefont {McEllistrem}, \citenamefont
  {Prados-Estévez}, \citenamefont {Vanhoy},\ and\ \citenamefont
  {Yates}}]{peters_level_2013}%
  \BibitemOpen
  \bibfield  {author} {\bibinfo {author} {\bibfnamefont {E.~E.}\ \bibnamefont
  {Peters}}, \bibinfo {author} {\bibfnamefont {A.}~\bibnamefont {Chakraborty}},
  \bibinfo {author} {\bibfnamefont {B.~P.}\ \bibnamefont {Crider}}, \bibinfo
  {author} {\bibfnamefont {B.~H.}\ \bibnamefont {Davis}}, \bibinfo {author}
  {\bibfnamefont {M.~K.}\ \bibnamefont {Gnanamani}}, \bibinfo {author}
  {\bibfnamefont {M.~T.}\ \bibnamefont {McEllistrem}}, \bibinfo {author}
  {\bibfnamefont {F.~M.}\ \bibnamefont {Prados-Estévez}}, \bibinfo {author}
  {\bibfnamefont {J.~R.}\ \bibnamefont {Vanhoy}},\ and\ \bibinfo {author}
  {\bibfnamefont {S.~W.}\ \bibnamefont {Yates}},\ }\bibfield  {title} {\bibinfo
  {title} {Level lifetimes in the stable {Zr} nuclei: {Effects} of chemical
  properties in {Doppler}-shift measurements},\ }\href
  {https://doi.org/10.1103/PhysRevC.88.024317} {\bibfield  {journal} {\bibinfo
  {journal} {Phys. Rev. C}\ }\textbf {\bibinfo {volume} {88}},\ \bibinfo
  {pages} {024317} (\bibinfo {year} {2013})}\BibitemShut {NoStop}%
\bibitem [{\citenamefont {Urban}\ \emph {et~al.}(2026)\citenamefont {Urban},
  \citenamefont {Abramuk}, \citenamefont {Rz{\c a}ca-Urban}, \citenamefont
  {Jentschel}, \citenamefont {Mutti}, \citenamefont {K{\"o}ster}, \citenamefont
  {de~France},\ and\ \citenamefont {Ur}}]{Urban}%
  \BibitemOpen
  \bibfield  {author} {\bibinfo {author} {\bibfnamefont {W.}~\bibnamefont
  {Urban}}, \bibinfo {author} {\bibfnamefont {A.}~\bibnamefont {Abramuk}},
  \bibinfo {author} {\bibfnamefont {T.}~\bibnamefont {Rz{\c a}ca-Urban}},
  \bibinfo {author} {\bibfnamefont {M.}~\bibnamefont {Jentschel}}, \bibinfo
  {author} {\bibfnamefont {P.}~\bibnamefont {Mutti}}, \bibinfo {author}
  {\bibfnamefont {U.}~\bibnamefont {K{\"o}ster}}, \bibinfo {author}
  {\bibfnamefont {G.}~\bibnamefont {de~France}},\ and\ \bibinfo {author}
  {\bibfnamefont {C.~A.}\ \bibnamefont {Ur}},\ }\href
  {https://arxiv.org/abs/2607.11409} {\bibinfo {title} {Structure of even-even
  {Zr} isotopes with $52\le {N} \le 58$ neutrons.}} (\bibinfo {year} {2026}),\
  \Eprint {https://arxiv.org/abs/2607.11409} {arXiv:2607.11409 [nucl-ex]}
  \BibitemShut {NoStop}%
\bibitem [{\citenamefont {Wood}\ \emph {et~al.}(1999)\citenamefont {Wood},
  \citenamefont {Zganjar}, \citenamefont {{De Coster}},\ and\ \citenamefont
  {Heyde}}]{Wood_E0}%
  \BibitemOpen
  \bibfield  {author} {\bibinfo {author} {\bibfnamefont {J.~L.}\ \bibnamefont
  {Wood}}, \bibinfo {author} {\bibfnamefont {E.~F.}\ \bibnamefont {Zganjar}},
  \bibinfo {author} {\bibfnamefont {C.}~\bibnamefont {{De Coster}}},\ and\
  \bibinfo {author} {\bibfnamefont {K.}~\bibnamefont {Heyde}},\ }\bibfield
  {title} {\bibinfo {title} {Electric monopole transitions from low energy
  excitations in nuclei},\ }\href
  {https://doi.org/https://doi.org/10.1016/S0375-9474(99)00143-8} {\bibfield
  {journal} {\bibinfo  {journal} {Nucl. Phys. A}\ }\textbf {\bibinfo {volume}
  {651}},\ \bibinfo {pages} {323} (\bibinfo {year} {1999})}\BibitemShut
  {NoStop}%
\bibitem [{\citenamefont {Werner}\ \emph {et~al.}(2008)\citenamefont {Werner},
  \citenamefont {Benczer-Koller}, \citenamefont {Kumbartzki}, \citenamefont
  {Holt}, \citenamefont {Boutachkov}, \citenamefont {Stefanova}, \citenamefont
  {Perry}, \citenamefont {Pietralla}, \citenamefont {Ai}, \citenamefont
  {Aleksandrova}, \citenamefont {Anderson}, \citenamefont {Cakirli},
  \citenamefont {Casperson}, \citenamefont {Casten}, \citenamefont
  {Chamberlain}, \citenamefont {Copos}, \citenamefont {Darakchieva},
  \citenamefont {Eckel}, \citenamefont {Evtimova}, \citenamefont {Fitzpatrick},
  \citenamefont {Garnsworthy}, \citenamefont {G\"urdal}, \citenamefont {Heinz},
  \citenamefont {Kovacheva}, \citenamefont {Lambie-Hanson}, \citenamefont
  {Liang}, \citenamefont {Manchev}, \citenamefont {McCutchan}, \citenamefont
  {Meyer}, \citenamefont {Qian}, \citenamefont {Schmidt}, \citenamefont
  {Thompson}, \citenamefont {Williams},\ and\ \citenamefont
  {Winkler}}]{Werner_gfact}%
  \BibitemOpen
  \bibfield  {author} {\bibinfo {author} {\bibfnamefont {V.}~\bibnamefont
  {Werner}}, \bibinfo {author} {\bibfnamefont {N.}~\bibnamefont
  {Benczer-Koller}}, \bibinfo {author} {\bibfnamefont {G.}~\bibnamefont
  {Kumbartzki}}, \bibinfo {author} {\bibfnamefont {J.~D.}\ \bibnamefont
  {Holt}}, \bibinfo {author} {\bibfnamefont {P.}~\bibnamefont {Boutachkov}},
  \bibinfo {author} {\bibfnamefont {E.}~\bibnamefont {Stefanova}}, \bibinfo
  {author} {\bibfnamefont {M.}~\bibnamefont {Perry}}, \bibinfo {author}
  {\bibfnamefont {N.}~\bibnamefont {Pietralla}}, \bibinfo {author}
  {\bibfnamefont {H.}~\bibnamefont {Ai}}, \bibinfo {author} {\bibfnamefont
  {K.}~\bibnamefont {Aleksandrova}}, \bibinfo {author} {\bibfnamefont
  {G.}~\bibnamefont {Anderson}}, \bibinfo {author} {\bibfnamefont {R.~B.}\
  \bibnamefont {Cakirli}}, \bibinfo {author} {\bibfnamefont {R.~J.}\
  \bibnamefont {Casperson}}, \bibinfo {author} {\bibfnamefont {R.~F.}\
  \bibnamefont {Casten}}, \bibinfo {author} {\bibfnamefont {M.}~\bibnamefont
  {Chamberlain}}, \bibinfo {author} {\bibfnamefont {C.}~\bibnamefont {Copos}},
  \bibinfo {author} {\bibfnamefont {B.}~\bibnamefont {Darakchieva}}, \bibinfo
  {author} {\bibfnamefont {S.}~\bibnamefont {Eckel}}, \bibinfo {author}
  {\bibfnamefont {M.}~\bibnamefont {Evtimova}}, \bibinfo {author}
  {\bibfnamefont {C.~R.}\ \bibnamefont {Fitzpatrick}}, \bibinfo {author}
  {\bibfnamefont {A.~B.}\ \bibnamefont {Garnsworthy}}, \bibinfo {author}
  {\bibfnamefont {G.}~\bibnamefont {G\"urdal}}, \bibinfo {author}
  {\bibfnamefont {A.}~\bibnamefont {Heinz}}, \bibinfo {author} {\bibfnamefont
  {D.}~\bibnamefont {Kovacheva}}, \bibinfo {author} {\bibfnamefont
  {C.}~\bibnamefont {Lambie-Hanson}}, \bibinfo {author} {\bibfnamefont
  {X.}~\bibnamefont {Liang}}, \bibinfo {author} {\bibfnamefont
  {P.}~\bibnamefont {Manchev}}, \bibinfo {author} {\bibfnamefont {E.~A.}\
  \bibnamefont {McCutchan}}, \bibinfo {author} {\bibfnamefont {D.~A.}\
  \bibnamefont {Meyer}}, \bibinfo {author} {\bibfnamefont {J.}~\bibnamefont
  {Qian}}, \bibinfo {author} {\bibfnamefont {A.}~\bibnamefont {Schmidt}},
  \bibinfo {author} {\bibfnamefont {N.~J.}\ \bibnamefont {Thompson}}, \bibinfo
  {author} {\bibfnamefont {E.}~\bibnamefont {Williams}},\ and\ \bibinfo
  {author} {\bibfnamefont {R.}~\bibnamefont {Winkler}},\ }\bibfield  {title}
  {\bibinfo {title} {Evidence for the microscopic formation of mixed-symmetry
  states from magnetic moment measurements},\ }\href
  {https://doi.org/10.1103/PhysRevC.78.031301} {\bibfield  {journal} {\bibinfo
  {journal} {Phys. Rev. C}\ }\textbf {\bibinfo {volume} {78}},\ \bibinfo
  {pages} {031301(R)} (\bibinfo {year} {2008})}\BibitemShut {NoStop}%
\bibitem [{\citenamefont {Flynn}\ \emph {et~al.}(1974)\citenamefont {Flynn},
  \citenamefont {Beery},\ and\ \citenamefont {Blair}}]{Flynn_tp}%
  \BibitemOpen
  \bibfield  {author} {\bibinfo {author} {\bibfnamefont {E.~R.}\ \bibnamefont
  {Flynn}}, \bibinfo {author} {\bibfnamefont {J.~G.}\ \bibnamefont {Beery}},\
  and\ \bibinfo {author} {\bibfnamefont {A.~G.}\ \bibnamefont {Blair}},\
  }\bibfield  {title} {\bibinfo {title} {The (t,p) reaction to the low-lying
  levels of the zirconium isotopes},\ }\href
  {https://doi.org/https://doi.org/10.1016/0375-9474(74)90005-0} {\bibfield
  {journal} {\bibinfo  {journal} {Nucl. Phys. A}\ }\textbf {\bibinfo {volume}
  {218}},\ \bibinfo {pages} {285} (\bibinfo {year} {1974})}\BibitemShut
  {NoStop}%
\bibitem [{\citenamefont {Ball}\ \emph {et~al.}(1971)\citenamefont {Ball},
  \citenamefont {Auble},\ and\ \citenamefont {Roos}}]{Ball_pt}%
  \BibitemOpen
  \bibfield  {author} {\bibinfo {author} {\bibfnamefont {J.~B.}\ \bibnamefont
  {Ball}}, \bibinfo {author} {\bibfnamefont {R.~L.}\ \bibnamefont {Auble}},\
  and\ \bibinfo {author} {\bibfnamefont {P.~G.}\ \bibnamefont {Roos}},\
  }\bibfield  {title} {\bibinfo {title} {Study of the {Z}irconium {I}sotopes
  with the ($p,t$) reaction},\ }\href {https://doi.org/10.1103/PhysRevC.4.196}
  {\bibfield  {journal} {\bibinfo  {journal} {Phys. Rev. C}\ }\textbf {\bibinfo
  {volume} {4}},\ \bibinfo {pages} {196} (\bibinfo {year} {1971})}\BibitemShut
  {NoStop}%
\bibitem [{\citenamefont {Tickle}\ \emph {et~al.}(1982)\citenamefont {Tickle},
  \citenamefont {Gray},\ and\ \citenamefont {Bent}}]{Tickle_Zr}%
  \BibitemOpen
  \bibfield  {author} {\bibinfo {author} {\bibfnamefont {R.~S.}\ \bibnamefont
  {Tickle}}, \bibinfo {author} {\bibfnamefont {W.~S.}\ \bibnamefont {Gray}},\
  and\ \bibinfo {author} {\bibfnamefont {R.~D.}\ \bibnamefont {Bent}},\
  }\bibfield  {title} {\bibinfo {title} {Two-proton pickup in the zirconium
  region using the ($^6${L}i, $^8${B}) reaction},\ }\href
  {https://doi.org/https://doi.org/10.1016/0375-9474(82)90067-7} {\bibfield
  {journal} {\bibinfo  {journal} {Nucl. Phys. A}\ }\textbf {\bibinfo {volume}
  {376}},\ \bibinfo {pages} {309} (\bibinfo {year} {1982})}\BibitemShut
  {NoStop}%
\bibitem [{\citenamefont {{Van Den Berg}}\ \emph {et~al.}(1984)\citenamefont
  {{Van Den Berg}}, \citenamefont {Saha}, \citenamefont {Jones}, \citenamefont
  {Put},\ and\ \citenamefont {Siemssen}}]{VanDenBerg}%
  \BibitemOpen
  \bibfield  {author} {\bibinfo {author} {\bibfnamefont {A.~M.}\ \bibnamefont
  {{Van Den Berg}}}, \bibinfo {author} {\bibfnamefont {A.}~\bibnamefont
  {Saha}}, \bibinfo {author} {\bibfnamefont {G.~D.}\ \bibnamefont {Jones}},
  \bibinfo {author} {\bibfnamefont {L.~W.}\ \bibnamefont {Put}},\ and\ \bibinfo
  {author} {\bibfnamefont {R.~H.}\ \bibnamefont {Siemssen}},\ }\bibfield
  {title} {\bibinfo {title} {Study of even-{A} zirconium and strontium isotopes
  with the ($d$,$^6${L}i) reaction},\ }\href
  {https://doi.org/https://doi.org/10.1016/0375-9474(84)90147-7} {\bibfield
  {journal} {\bibinfo  {journal} {Nucl. Phys. A}\ }\textbf {\bibinfo {volume}
  {429}},\ \bibinfo {pages} {1} (\bibinfo {year} {1984})}\BibitemShut {NoStop}%
\bibitem [{\citenamefont {Kib\'edi}\ \emph {et~al.}(2022)\citenamefont
  {Kib\'edi}, \citenamefont {Garnsworthy},\ and\ \citenamefont
  {Wood}}]{KibediE0}%
  \BibitemOpen
  \bibfield  {author} {\bibinfo {author} {\bibfnamefont {T.}~\bibnamefont
  {Kib\'edi}}, \bibinfo {author} {\bibfnamefont {A.~B.}\ \bibnamefont
  {Garnsworthy}},\ and\ \bibinfo {author} {\bibfnamefont {J.~L.}\ \bibnamefont
  {Wood}},\ }\bibfield  {title} {\bibinfo {title} {Electric monopole
  transitions in nuclei},\ }\href
  {https://doi.org/https://doi.org/10.1016/j.ppnp.2021.103930} {\bibfield
  {journal} {\bibinfo  {journal} {Prog. Part. Nucl. Phys.}\ }\textbf {\bibinfo
  {volume} {123}},\ \bibinfo {pages} {103930} (\bibinfo {year}
  {2022})}\BibitemShut {NoStop}%
\bibitem [{\citenamefont {Leoni}\ \emph {et~al.}(2024)\citenamefont {Leoni},
  \citenamefont {Fornal}, \citenamefont {Bracco}, \citenamefont {Tsunoda},\
  and\ \citenamefont {Otsuka}}]{Leoni_PPNP}%
  \BibitemOpen
  \bibfield  {author} {\bibinfo {author} {\bibfnamefont {S.}~\bibnamefont
  {Leoni}}, \bibinfo {author} {\bibfnamefont {B.}~\bibnamefont {Fornal}},
  \bibinfo {author} {\bibfnamefont {A.}~\bibnamefont {Bracco}}, \bibinfo
  {author} {\bibfnamefont {Y.}~\bibnamefont {Tsunoda}},\ and\ \bibinfo {author}
  {\bibfnamefont {T.}~\bibnamefont {Otsuka}},\ }\bibfield  {title} {\bibinfo
  {title} {Multifaceted character of shape coexistence phenomena in atomic
  nuclei},\ }\href {https://doi.org/https://doi.org/10.1016/j.ppnp.2024.104119}
  {\bibfield  {journal} {\bibinfo  {journal} {Prog. Part. Nucl. Phys.}\
  }\textbf {\bibinfo {volume} {139}},\ \bibinfo {pages} {104119} (\bibinfo
  {year} {2024})}\BibitemShut {NoStop}%
\end{thebibliography}
\end{document}